\documentclass[floats,floatfix,amssymb,prd,twocolumn,nofootinbib,superscriptaddress,
]{revtex4-2}

\usepackage[utf8]{inputenc}
\usepackage{amsmath,amssymb}
\usepackage{amsfonts}
\usepackage{bm}
\usepackage{courier}
\usepackage{graphics,orcidlink}
\usepackage{float}
\usepackage{varioref}
\usepackage{mathtools}
\usepackage[normalem]{ulem} 
\usepackage{acronym}
\usepackage{float}
\usepackage[caption=false]{subfig}
\usepackage{lipsum}
\usepackage{multirow}
\usepackage{graphicx}
\usepackage{tensor}
\usepackage{verbatim}
\usepackage{wrapfig}
\usepackage{soul} 
\allowdisplaybreaks

\labelformat{section}{Section #1} 
\labelformat{subsubsection}{Section #1}
\labelformat{subsubsubsection}{Section #1}
\labelformat{equation}{Eq.~(#1)} 
\labelformat{figure}{Fig.~#1} 
\labelformat{subfigure}{Fig.~\thefigure#1} 
\labelformat{table}{Tab.~#1} 
\labelformat{appendix}{Appendix #1}

\begin{document}

\title{Dynamical tidal Love numbers of Kerr-like compact objects
}

\author{Sumanta Chakraborty\orcidlink{0000-0003-3343-3227}}
\email{tpsc@iacs.res.in}
\affiliation{School of Physical Sciences, Indian Association for the Cultivation of Science, Kolkata-700032, India}

\author{Elisa Maggio\orcidlink{0000-0002-1960-8185}}
\email{elisa.maggio@aei.mpg.de}
\affiliation{Max Planck Institute for Gravitational Physics (Albert Einstein Institute), D-14476 Potsdam, Germany}
\affiliation{Niels Bohr International Academy, Niels Bohr Institute, Blegdamsvej 17, 2100 Copenhagen, Denmark}

\author{Michela Silvestrini\orcidlink{0009-0004-2250-088X}}
\email{michela.silvestrini@unina.it}
\affiliation{Dipartimento di Fisica, ``Sapienza'' Universit\`a di Roma \& Sezione INFN Roma1, P.A. Moro 5, 00185, Roma, Italy}
\affiliation{Dipartimento di Fisica “E. Pancini”, Università degli studi di Napoli Federico II, Compl. Univ. di Monte S. Angelo, Via Cintia, I-80126 Napoli, Italy}
\affiliation{INAF -- Osservatorio Astronomico di Capodimonte, Salita Moiariello 16, I-80131, Napoli, Italy}

\author{Paolo Pani\orcidlink{0000-0003-4443-1761}}
\email{paolo.pani@uniroma1.it}
\affiliation{Dipartimento di Fisica, ``Sapienza'' Universit\`a di Roma \& Sezione INFN Roma1, P.A. Moro 5, 00185, Roma, Italy}

\begin{abstract}
We develop a framework to compute the tidal response of a Kerr-like compact object in terms of its reflectivity, compactness, and spin, both in the static and the frequency-dependent case. Here we focus on the low-frequency regime, which can be solved fully analytically.
We highlight some remarkable novel features, in particular: i) Even in the zero-frequency limit, the tidal Love numbers~(TLNs) depend on the linear-in-frequency dependence of the object's reflectivity in a nontrivial way. ii) {Intriguingly, although the static limit of the (phenomenologically more interesting) frequency-dependent TLNs is continuous,
it differs from the strictly static TLNs for compact objects other than black holes}.
This shows that earlier findings regarding the static TLNs of ultracompact objects correspond to a measure-zero region in the parameter space, though the logarithmic behavior of the TLNs in the black hole limit is retained. 
iii) In the non-rotating case, the TLNs \emph{generically vanish} in the zero-frequency limit (just like for a black hole), except when the reflectivity is ${\cal R}=1+\,{\cal O}(M\omega)$, in which case they vanish with a model-dependent scaling, which is generically logarithmic, in the black-hole limit.
The TLNs initially grow with frequency, for any nonzero reflectivity, and then display oscillations and resonances tied up with the quasi-normal modes of the object.
iv) For rotating compact objects, the TLNs decrease when the reflectivity decreases or the rotation parameter increases. 
Our results lay the theoretical groundwork to develop model-independent tests of the nature of compact objects using tidal effects in gravitational-wave signals.
\end{abstract}
\maketitle
\section{Introduction}

According to the theory of general relativity~(GR), a black hole~(BH) has vanishing tidal susceptibility. Namely, when immersed in a quasi-stationary tidal field (for example the one produced by a binary companion at a large distance) its induced multipole moments are all zero.
This special property is quantified by the vanishing of the static tidal Love numbers~(TLNs)~\cite{Damour_tidal, Binnington:2009bb, Damour:2009vw, Gurlebeck:2015xpa, Poisson:2014gka, Pani:2015hfa, Landry:2015zfa, LeTiec:2020bos, Chia:2020yla, LeTiec:2020spy} and can be related to some ``hidden" symmetries of the Kerr solution in GR~\cite{Hui:2020xxx, Charalambous:2021kcz, Charalambous:2021mea, Hui:2021vcv, Berens:2022ebl, BenAchour:2022uqo, Charalambous:2022rre, Katagiri:2022vyz, Ivanov:2022qqt, DeLuca:2023mio}. It is a quite unique and fragile feature~\cite{Porto:2016zng, Cardoso:2017cfl}: it does not hold true for any other object, even within GR~\cite{Cardoso:2017cfl, Sennett:2017etc, Mendes:2016vdr, Pani:2015tga, Cardoso:2017cfl, Uchikata:2016qku, Raposo:2018rjn, Cardoso:2019rvt}, for BHs in modified gravity~\cite{Cardoso:2017cfl, Cardoso:2018ptl}, or in higher dimensions~\cite{Chakravarti:2018vlt, Chakravarti:2019aup, Cardoso:2019vof, Pereniguez:2021xcj,Charalambous:2023jgq, Rodriguez:2023xjd, Dey:2020lhq, Dey:2020pth}.
Its violation can therefore be seen as a smoking gun for deviations from the standard paradigm involving BHs in GR~\cite{Cardoso:2017cfl, Cardoso:2019rvt}, especially for supermassive compact objects~\cite{Maselli:2018fay, Datta:2021hvm, Pani:2019cyc, Piovano:2022ojl} which, in the standard paradigm, can only be BHs. So far the TLNs of compact objects other than Kerr BHs have been computed on a case-by-case basis, including boson stars~\cite{Cardoso:2017cfl, Sennett:2017etc, Mendes:2016vdr}, gravastars~\cite{Pani:2015tga, Cardoso:2017cfl, Uchikata:2016qku}, anisotropic stars~\cite{Raposo:2018rjn}, and other simple exotic compact objects~(ECOs)~\cite{Giudice:2016zpa, Cardoso:2019rvt} with stiff equation of state at the surface~\cite{Cardoso:2017cfl}. Furthermore, they have been mostly computed for nonspinning objects and for static tidal fields, with the notable exceptions of Refs.~\cite{Chia:2020yla,Charalambous:2021mea,Bonelli:2021uvf,Creci:2021rkz, Consoli:2022eey, Saketh:2023bul,Perry:2023wmm}.

In the context of gravitational-wave tests of the nature of compact objects based on their tidal response (see~\cite{Cardoso:2019rvt, Maggio:2021ans} for some reviews), it would be useful to have a framework to compute the TLNs from the general properties of the object and not restricted to some specific underlying model.
The scope of this work is to provide such a general framework for the TLNs of a Kerr-like compact object, by carefully taking into account the reflectivity of the object, the role of the angular momentum, and the possible time dependence of the tidal field.
We exploit a better understanding of the role of the reflectivity of ECOs, recently investigated in the context of their quasinormal modes~(QNMs) in the ringdown~\cite{Maggio:2018ivz, Maggio:2020jml}. While our framework is valid for any frequency, here we focus on the low-frequency regime which is amenable to analytical treatment using matching asymptotics.

In~\ref{sec:setup} we present our setup, which is based on considering a Kerr metric with generic boundary conditions at a finite distance from the would-be horizon and assuming GR (at least) in the exterior of the compact object. We use the formalism developed in Ref.~\cite{Chia:2020yla} (see also~\cite{Consoli:2022eey, Bhatt:2023zsy}) to derive a master equation for the TLNs, paying particular attention to the object's reflectivity, ${\cal R}$. We emphasize that the reflectivity of any compact object is more naturally defined with respect to plane wave modes near the horizon. It is therefore convenient to define the reflectivity in terms of a master function describing linear perturbations that are wave-like near the horizon. For this reason, we use the Detweiler function rather than the more standard Teukolsky function, following closely Ref.~\cite{Maggio:2020jml}. 

Another important aspect of our analysis is the static limit, particularly for the non-rotating case. As we will demonstrate, quite counter-intuitively, the $\omega\rightarrow 0$ limit of the frequency-dependent TLNs for a non-rotating compact object with $\mathcal{R} \neq 0$ does not coincide with the strictly static TLNs derived in~\cite{Cardoso:2017cfl}, though the frequency-dependent TLNs are continuous throughout the low-frequency regime. 
The same behavior was observed in~\cite{Pani:2018inf} in the context of the magnetic TLNs of fluid stars. In particular, it was found that the $\omega \to 0$ limit of the dynamical TLNs for neutron stars does not coincide with the strictly static TLNs. This difference is due to the existence of two types of magnetic TLNs (static and irrotational), which nicely map to the existing results in the literature~\cite{Binnington:2009bb,Damour:2009vw}.

In our context, the difference between the strictly static TLNs and their $\omega\rightarrow 0$ limit happens because the solutions of the Teukolsky equation for an ECO, admitting outgoing behavior near the horizon, with $\omega \neq 0$, do not tend to the respective solutions in the zero-frequency limit\footnote{{Note that in the BH limit only the ingoing solution to the Teukolsky equation matters, which has a continuous zero-frequency limit. Thus, for BHs the dynamical TLNs in the $\omega \to 0$ limit coincides with the strictly static TLNs. The difference between the $\omega \to 0$ and the strictly static scenario arises from the reflective part associated with the out-going solution, present for both ECOs and NSs. This issue was also discussed in \cite{LeTiec:2020bos}, pointing out that the two independent solutions of the Teukolsky equation with $\omega \neq 0$, do not remain independent in the $\omega \to 0$ limit.}}. Hence the strictly static TLNs of a non-rotating compact object must be determined in a separate manner and do not follow from a continuous $\omega \rightarrow 0$ limit. 

Another interesting result is that the zero-frequency limit of the frequency-dependent TLNs depends on the small-frequency limit of the reflectivity, which we model as
\begin{equation}
{\cal R}(\omega) = {\cal R}_0+ iM\omega {\cal R}_1 +{\cal O}(M^2\omega^2) \,,
\label{reflectivity}
\end{equation}
where $\omega$ is the frequency of the tidal perturbation, $M$ is the object's mass, and the (complex) coefficients ${\cal R}_i$ generically depend on the object's properties. 
One of our main findings is that, due to some cancellations, the coefficient ${\cal R}_1$ of the ${\cal O}(M\omega)$ term in the above equation enters the TLNs in the \emph{static} limit ($\omega\to0$).
We, therefore, argue that modeling the tidal deformability of a compact object requires knowledge of the frequency dependence of the reflectivity, even when restricting to the \emph{static} TLNs. Moreover, we find that, in the static limit and for non-rotating objects, the \emph{TLNs vanish identically} (as in the BH case), except when ${\cal R}_0=1$, in which case 
the corresponding nonzero value of the TLNs depends on ${\cal R}_1$ and exhibits a logarithmic behavior as a function of the compactness parameter as the BH limit is approached. A logarithmic dependence was also found in Ref.~\cite{Cardoso:2017cfl}
by strictly static perturbations of  perfectly reflecting, non-spinning objects. However, as we shall discuss, beside the logarithmic dependence the general expression of the TLN in the zero-frequency limit differs from the one found in Ref.~\cite{Cardoso:2017cfl} in the strictly static case. 
This again shows that a strictly static limit cannot be obtained in a continuous manner from the frequency-dependent TLNs.
Since the tidal field sourcing the tidal deformation is intrinsically dynamical in binary coalescences, the static TLNs should be computed as the $\omega\to 0$ limit of the dynamical ones (see~\cite{Landry:2015cva,Pani:2018inf} for a related discussion).
Intriguingly, the logarithmic behavior as a function of the compactness parameter for the TLNs of compact objects is a generic result, which holds in the strictly static case as well as in the dynamical contexts, but in the dynamical case it depends on the choice of $\mathcal{R}_{1}$ as well. In particular, it is possible to choose $\mathcal{R}_{1}$, such that the TLNs do not have a logarithmic dependence. 

In~\ref{sec:results} we present a collection of results for the TLNs of a compact object in terms of its reflectivity, compactness, and spin, both in the static and in the frequency-dependent case. For the case of a non-rotating compact object, we discuss the static limit in two different manners --- (a) substituting $\omega=0$ in the Teukolsky equation and then determining the (strictly static) TLNs from its solution, which coincides with the results of~\cite{Cardoso:2017cfl}; (b) solving the dynamical case and thus determining the dynamical TLNs, and then taking $\omega\rightarrow 0$ limit. 
As mentioned above, the static limit of the TLNs associated with a non-rotating compact object depends on $\mathcal{R}_{0}$ and $\mathcal{R}_{1}$, and the TLNs are either zero or vanish in the BH limit with a scaling that depends on the compactness and on the model-dependent choice of ${\cal R}_1$. Though the non-zero TLNs generically depend on the compactness in a logarithmic fashion, only when setting $\omega=0$ from the beginning, do we recover the results of the previous work~\cite{Cardoso:2017cfl}.

Finally, in the more general spinning and frequency-dependent case we determine the TLNs in the small frequency limit. We present the dependence of the TLNs of rotating compact objects on the model parameters, showing that the induced multipole moments decrease as the rotation increases or as the reflectivity decreases. Interestingly, the dynamical TLNs in the non-rotating case depict certain oscillatory and resonant behaviors which are seemingly characteristics of the low-frequency QNMs~\cite{Maggio:2017ivp, Maggio:2018ivz, Maggio:2020jml} associated with the compact object.

We conclude in~\ref{sec:conclusions}, with a discussion of our main findings and possible extensions. Several detailed calculations have been presented in the appendices for the benefit of the reader.
Throughout this work, we use geometrized $G=1=c$ units, unless otherwise mentioned. We also work with the signature convention such that flat Minkowski metric in the Cartesian coordinate system reads $\eta_{\mu \nu}=\textrm{diag.}(-1,+1,+1,+1)$. 
\section{Setup}\label{sec:setup}

\subsection{Exterior geometry of Kerr-like compact objects}\label{sec:ECO}

In this work, we analyze the tidal effects on a spinning and horizonless compact object whose exterior spacetime is described by the Kerr metric. This is an approximation since, beyond spherical symmetry, the external spacetime of a compact object can have arbitrary multipole moments, even in GR. At the same time, regularity requires that any multipolar deviation from the Kerr metric should vanish in the BH limit at least within GR~\cite{Raposo:2018xkf}.

It is useful to express the Kerr metric in terms of the ingoing null coordinates $( v, r, \theta, \tilde{\phi})$ (their relation to Boyer-Lindquist coordinates $(t,r,\theta,\phi)$ is given in~\ref{TeuDet})
such that the line element of the exterior Kerr spacetime reads
\begin{eqnarray}
\nonumber  
ds^2 &=& - \frac{\left( \Delta - a^2 \sin^2 \theta\right)}{\Sigma} dv^2 + 2 dv dr 
\\
 &-& 2 a \sin^2 \theta dr d\tilde{\phi} - \frac{4 M a r \sin^2 \theta}{\Sigma} dv d\tilde{\phi} 
 \\
\nonumber 
&+& {\Sigma}d\theta^2+\frac{\left[(r^2+a^2)^2 - \Delta a^2 \sin^2 \theta\right]}{\Sigma} \sin^2 \theta d\tilde{\phi}^2 \,,
\label{kerr}
\end{eqnarray}
where $\Sigma=r^2+a^2\cos^2\theta$, $\Delta=(r-r_+)(r-r_-)$, and $r_{\pm} = M \pm \sqrt{M^2 - a^2}$ are the locations of the horizons. The parameters $M$ and $J\equiv aM$ in the line element are the total mass and the angular momentum of the compact object, respectively. 

The radius $r_0$ of the compact object is assumed to be only slightly larger than the event horizon, located at $r=r_{+}$, and smaller than the photon sphere. Such a scenario can be motivated by models of microscopic corrections at the horizon scale (see~\cite{Cardoso:2019rvt} for an overview). For the purpose of this work, we consider the radius of the compact object to be
\begin{equation}\label{radius_compact}
r_0 = r_+ (1+\epsilon)\,,
\end{equation}
where $0 < \epsilon \ll 1$. The properties of the object's interior are parametrized by a complex and frequency-dependent reflectivity $\mathcal{R}(\omega)$ at the surface\footnote{Note that, within our formalism, $r_0$ can be any effective radius in the exterior geometry, as long as the reflectivity $\mathcal{R}(\omega)$ refers to the surface at $r=r_0$. For example, if the object deviates from Kerr only near the would-horizon, one could take $r_0$ to be sufficiently large so that the hypothesis of a Kerr background for $r>r_0$ is satisfied.} of the compact object, as will be discussed in detail in Section~\ref{sec:reflectivity}. 

\subsection{Tidal Love numbers~(TLNs)}\label{sec:perturbations}

Here, we wish to provide the basic framework for computing the deformation of a Kerr-like compact object when placed in an external tidal field. In particular, we wish to study the TLNs, which encapsulate the linear response of the compact object due to a perturbing tidal field. We first present the Newtonian description of the tidal response and then generalize it to the relativistic case. 

\subsubsection{Newtonian TLNs}

The study of the tidal response of a self-gravitating object due to an external tidal field was historically performed in the Newtonian framework, where the total gravitational potential outside the object in the external tidal field can be divided into two parts (see, e.g.,~\cite{PoissonWill}): (a) $U_{\rm tidal}$, the gravitational potential due to the tidal field, and (b) $U_{\rm body}$, the gravitational potential due to the compact object. The tidal potential can be decomposed in the spherical harmonic basis as $U_{\rm tidal}\sim \sum_{\ell m} \mathcal{E}_{\ell m}Y_{\ell m}r^{\ell}$, where $\mathcal{E}_{\ell m}$ are the components of the tidal field associated with the spherical harmonic decomposition, $\ell \geq 0$ is the angular number, and $m$ is the azimuthal number (with $|m| \leq \ell$). In general, $\mathcal{E}_{\ell m}$ will involve $\ell$ derivatives of the tidal potential. Similarly, the Newtonian potential due to the central object can be expanded in terms of its multipole moments, $I_{\ell m}$, as $U_{\rm body}\sim \sum_{\ell m} I_{\ell m}Y_{\ell m}r^{-\ell-1}$. At the linear level, the deviation $\delta I_{\ell m}$ from the background multipole moments (those of a Kerr metric~\cite{Hansen:1974zz} in our case) are proportional to the tidal field $\mathcal{E}_{\ell m}$ and its time derivatives,
\begin{align}
\delta I_{\ell m}\propto r_{0}^{2\ell+1}\left(2k_{\ell m}\mathcal{E}_{\ell m}-\tau_{0}\nu_{\ell m}\dot{\mathcal{E}}_{\ell m}+...\right)~.
\end{align}
Here, $k_{\ell m}$ are the TLNs (related to the conservative response), $\tau_{0}$ is a characteristic timescale associated with the tidal field, $\nu_{\ell m}$ is associated with the tidal dissipation of the compact object (sometimes called tidal heating in different contexts~\cite{Hartle:1973zz, Hughes:2001jr, Maselli:2017cmm, Datta:2019epe, Chakraborty:2021gdf}), and the dots denote higher-order derivatives with respect to the Boyer-Lindquist time coordinate. Note that the proportionality factor due to the linear relation between the multipole and tidal moments depends on the nature of the compact object and its size, namely on $r_{0}$. Given the above multipolar expansion, the total potential outside the compact object can be expressed as
\begin{align}
U_{\rm total}&=U_{\rm body}+U_{\rm tidal}=-\frac{M}{r}+{\rm intrinsic~moments}
\nonumber
\\
&+ \int d\omega \sum_{\ell m}\mathcal{E}_{\ell m}r^{\ell}\left[1+F_{\ell m}(\omega)\left(\frac{r_{0}}{r}\right)^{2\ell+1}\right]Y_{\ell m}~,
\label{Newton_pot}
\end{align}
where the frequency-dependent response function $F_{\ell m}(\omega)$ reads
\begin{align}
F_{\ell m}(\omega)=2k_{\ell m}(\omega)+i\omega \tau_{0}\nu_{\ell m}(\omega) \,,
\end{align}
and the (conservative) Newtonian TLNs are given by $k_{\ell m}=(1/2)\textrm{Re}F_{\ell m}$. 

\subsubsection{Relativistic TLNs}\label{sec:relTLN}

A gauge-invariant, relativistic extension of the Newtonian TLNs can be achieved with the use of Weyl scalars in the Newman-Penrose formalism~\cite{Newman:1961qr}. The Weyl scalar which describes the behavior of the gravitational waves that propagate towards asymptotic infinity is $\Psi_{4}$ and hence is the one we will use in this work. In the Kerr-like background geometry, the Weyl scalar $\Psi_{4}$ can be decomposed as~\cite{Teukolsky:1972my}
\begin{equation}\label{psi4}
\rho^4 \Psi_4 =\int d\omega\sum_{\ell m} R_{\ell m}(r) ~_{-2}S_{\ell m}(\theta) e^{i (m \phi -\omega v)} \,,
\end{equation}
where $\rho = -(r-ia \cos \theta)^{-1}$, $R_{\ell m}(r)$ is the radial Teukolsky function, and $_{-2}S_{\ell m}(\theta) e^{i m \phi}$ is the spin-weighted spheroidal harmonic with spin $s=-2$.  

The fact that $\Psi_{4}$ is able to describe the tidal effects can be seen by computing it in the Newtonian limit. In this case, the only nontrivial metric functions read $g_{00}=-(1+2c^{-2}U_{\rm total})$ and $g_{ij}=(1+2c^{-2}U_{\rm total})\delta_{ij}$, where $U_{\rm total}$ is the Newtonian potential including the tidal effects and we have restored the speed of light for the moment. In the Newtonian limit, obtained as $c\to\infty$, the computation of $\Psi_{4}$ reveals the following behavior~\cite{Chia:2020yla}:
\begin{align}\label{limpsi4}
&\lim_{c\rightarrow \infty}c^{2}\Psi_{4}\propto \nonumber \\
&\int d\omega \sum_{\ell m}\mathcal{E}_{\ell m}r^{\ell-2}\left[1+F_{\ell m}(\omega) \left(\frac{r_{0}}{r}\right)^{2\ell+1}\right]~_{-2}Y_{\ell m}\,,
\end{align}
where $_{-2}Y_{\ell m}$ are the spin-weighted spherical harmonics. The above behavior is valid in the intermediate region $r_0\ll r\ll 1/\omega$, i.e. sufficiently far away both from the central object and the tidal source.
As evident, the radial part of the above equation is identical to the one derived from the Newtonian potential in~\ref{Newton_pot}, with $F_{\ell m}$ being the tidal response of the compact object to the external tidal field, modulo the external power of the radial coordinate $r$. In \ref{Newton_pot}, the external radial dependence scales as $r^{\ell}$, while in \ref{limpsi4} the radial dependence behaves as $r^{\ell-2}$. This difference can be accounted for by noting that in the Newtonian limit $\Psi_{4}\sim \bar{m}^{i}\bar{m}^{j}\nabla_{i}\nabla_{j}U_{\rm total}$, where $\bar{m}^{i}$ is one of the complex vector in the Newman-Penrose formalism, $\nabla_{i}$ is the three-dimensional covariant derivative, and $U_{\rm total}$ has been defined earlier. The null vector $\bar{m}^{i}$ has a radial dependence $\sim (1/r)$ and is the origin of the $(1/r^{2})$ difference between the Newtonian potential and the Weyl scalar.  

The radial dependence in~\ref{limpsi4} is valid also in the relativistic case in the intermediate region $r_0\ll r\ll 1/\omega$. In this region, we can extract the tidal response function $F_{\ell m}(\omega)$ from $\Psi_4$ and define the relativistic TLNs in a gauge invariant manner as 
\begin{align}
k_{\ell m}=\frac{1}{2}\text{Re}F_{\ell m}~. \label{TLNs}
\end{align}

In general, the determination of the TLNs from $\Psi_{4}$ is achieved through the following procedure: (i) find $\Psi_{4}$ in the near-horizon region and apply the appropriate boundary conditions at $r_0$ (which will be discussed in Section~\ref{sec:reflectivity}), (ii) determine $\Psi_{4}$ in an intermediate region, which is far away from both the compact object and the source of the tidal field, (iii) for each $(\ell,m)$, determine the tidal response as the coefficient of $r^{-\ell-3}$ in~\ref{limpsi4}, (iv) determine the real part of the tidal response, which provides the corresponding TLN through~\ref{TLNs}. 
In the following, we perform each of these steps and provide the basic set-up for obtaining the TLNs. For brevity, we shall omit the $(\ell,m)$ subscripts in the radial and angular functions in the Weyl scalar. 

The radial function $R(r)$ satisfies the (source-free) radial Teukolsky equation~\cite{Teukolsky:1972my,Teukolsky:1973ha,Teukolsky:1974yv}, which in the Kinnersley null tetrad~\cite{doi:10.1063/1.1664958} and in the ingoing null coordinate system, reads
\begin{eqnarray}
\nonumber &&\frac{d^2 R(r)}{dr^2} + \left( \frac{2 i P_+ - 1}{r-r_+} - \frac{2 i P_- + 1}{r-r_-} - 2 i \omega \right) \frac{d R(r)}{dr} 
\\
\nonumber 
&&+ \left( \frac{4 i P_-}{(r-r_-)^2} - \frac{4 i P_+}{(r-r_+)^2} + \frac{A_- + i B_-}{(r-r_-)(r_+ - r_-)} \right. 
\\
&&\qquad - \left. \frac{A_+ + i B_+}{(r-r_+)(r_+ - r_-)} \right) R(r) =0 \,, \label{Teukeq}
\end{eqnarray}
where the quantities $P_{\pm}$, $A_{\pm}$ and $B_{\pm}$ are defined as
\begin{eqnarray}
    P_{\pm} &=& \frac{a m - 2 r_{\pm} M \omega}{r_+ - r_-} \,, 
    \label{Pplusminus}
    \\
    \nonumber A_{\pm} &=& E_{\ell m} - 2 - 2 (r_+ - r_-) P_{\pm} \omega \\
    &&- (r_{\pm} + 2 M) r_{\pm} \omega^2 \,, \\
    B_{\pm} &=& 2 r_{\pm} \omega\,, 
\end{eqnarray}
and $E_{\ell m}=\ell(\ell+1)+\mathcal{O}(a\omega)$. We would like to emphasize that the structure of the above differential equation is different in the static limit of the non-rotating case, i.e., for $a=0=\omega$. This will have important implications for the static limit and for the dependence of the TLNs on the compactness parameter $\epsilon$, which we will discuss in a later section. 

An analytical expression for the TLNs can be derived from the small-frequency approximation of the radial Teukolsky equation in~\ref{Teukeq}. The details of the small-frequency expansion can be found in~\ref{app:smallfrequency}. The solution of the Teukolsky equation is obtained in the small-frequency (i.e., $M\omega\ll1$) and near-zone (i.e., $r\omega\ll1$) approximation. The near-zone region includes both the near-horizon ($r-r_{+}\ll r_{+}$) and intermediate ($r_{+}\ll r\ll 1/\omega$) regions. 
Hence, we first apply the appropriate boundary condition to the near-horizon limit of the near-zone solution, and then obtain the TLNs in the intermediate zone.
The near-horizon limit of the solution yields (see~\ref{app:smallfrequency} for a derivation),
\begin{align}\label{near_horizon}
%
R(r) \sim \frac{\mathcal{A}}{(r_{+}-r_{-})^{4}}\Delta^{2}+\mathcal{B}e^{2i\bar{\omega}r_{*}}~, \quad r \to r_+~,
\end{align}
where, $\bar{\omega}=\omega-m\Omega_{\rm H}$, $\Omega_{\rm H}$ being the angular velocity of the horizon, and $r_{*}$ is the tortoise coordinate defined as $dr_{*}/dr=\left(r^{2}+a^{2}\right)/\Delta$. The unknown functions $\mathcal{A}$ and $\mathcal{B}$ depend on the frequency and are to be determined from the boundary conditions, which will be discussed in Section~\ref{sec:reflectivity}. After imposing these boundary conditions, the Teukolsky equation can be solved up to the intermediate region, far from both the compact object and the source of the tidal field, yielding the response function of the compact object to the external tidal field. Finally, the TLN of the $(\ell,m)$ mode reads in the small-frequency approximation (see~\ref{app:smallfrequency} for a derivation),
\begin{align}\label{tidal_love_small}
k_{\ell m}=\textrm{Re}\Bigg[-i\frac{P_{+}}{2}\left(\frac{(\ell+2)!(\ell-2)!}{(2\ell)!(1+2\ell)!}\right)&\prod_{j=1}^{\ell}\left(j^{2}+4P_{+}^{2}\right)
\nonumber
\\
\qquad \times\left\{\frac{1-\frac{\mathcal{B}}{\mathcal{A}}\Gamma_{1}}{1+\frac{\mathcal{B}}{\mathcal{A}}\Gamma_{1}}\right\}&\Bigg]~,
\end{align}
where,
\begin{align}
\Gamma_{1}=\frac{(\ell+2)!}{(\ell-2)!}\frac{\left(3+2iP_{+}\right)_{\ell-2}}{\left(-1-2iP_{+}\right)_{\ell+2}}~.
\end{align}
In the expression for $\Gamma_{1}$, we have introduced the Pochhammer symbol, $z_{n}\equiv z(z+1)\times \cdots \times (z+n-1)$. A similar expression (related to the imaginary part of $F_{\ell m}$) gives the dissipative response $\nu_{\ell m}$.

As expected, the tidal response depends on the ratio ${\cal B}/{\cal A}$ and, for $\mathcal{B}=0$ (i.e., in the BH limit), the response function becomes imaginary, leading to vanishing TLNs. However, for compact objects other than BHs, $\mathcal{B}$ is non-zero, and hence it is uniquely determined by the boundary conditions at $r=r_0$, which we now discuss. 

\subsection{Reflectivity of compact objects}\label{sec:reflectivity}

As shown in Section~\ref{sec:perturbations}, the TLNs of Kerr-like compact objects computed in~\ref{tidal_love_small} depend on the unknown constants of integration, $\mathcal{A}$ and $\mathcal{B}$, whose ratio is fixed by the boundary conditions at $r=r_0$. The radial Teukolsky function in the near horizon regime has two contributions~\cite{Teukolsky:1974yv}: (a) an ingoing part, which behaves as $\Delta^{2}$ and (b) an outgoing part, with radial dependence $e^{2i\bar{\omega}r_{*}}$. Thus, from~\ref{near_horizon}, one may identify the constant $\mathcal{A}$ to be associated with the ingoing mode at the horizon, while the constant $\mathcal{B}$ is associated with the outgoing mode at the horizon. For a BH, regularity of the perturbation at $r\sim r_+$ imposes no outgoing modes at the horizon since the BH absorbs everything and hence one should set $\mathcal{B}=0$. 

For non-BH objects, be it a neutron star, an ECO~\cite{Giudice:2016zpa, Cardoso:2019rvt, Maggio:2021ans} or a quantum corrected BH~\cite{Oshita:2019sat, Chakraborty:2022zlq, Nair:2022xfm}, there would be nontrivial reflection by the object's interior, and hence there would be outgoing modes for which $\mathcal{B}\neq 0$. Overall, the properties of the compact object are embedded into a single quantity, defined as the reflectivity of the object. BHs have zero reflectivity, ordinary neutron stars have almost perfect reflectivity~\cite{Maggio:2018ivz} (due to viscosity they have tiny interaction with gravitational waves, but see~\cite{Ripley:2023qxo} for a recent analysis of the out-of-equilibrium tidal dynamics), while non-BH compact objects will in general have nonzero and frequency-dependent reflectivity. We parametrize the latter by a complex number ${\cal R}(\omega)$, which might generically introduce a phase shift in the reflected wave relative to the incident one.

Loosely speaking, the reflectivity is the ratio of the amplitudes of the outgoing and the ingoing waves and one may consider the ratio $(\mathcal{B}/\mathcal{A})$ as the reflectivity. However, this would not be a correct assessment, since the outgoing and ingoing modes in terms of the radial Teukolsky function in the $(v,r,\theta,\tilde{\phi})$ coordinate are \emph{not} plane waves and hence the energy carried by ingoing and outgoing modes is not  $|{\cal A}|^2$ and $|{\cal B}|^2$, but rather a more complicated function of the frequency~\cite{Teukolsky:1974yv}. The way around this difficulty is to express the radial Teukolsky equation as a Schr\"{o}dinger-like equation with a real potential, which is constant in the near horizon regime, in which case the mode functions near the horizon are exact plane waves. This is achieved by the Detweiler function, $X$, which is defined as a linear combination of the radial Teukolsky function $R^{(t)}$ in the Boyer-Lindquist coordinates and its derivative as~\cite{Detweiler:1977, Maggio:2018ivz} 
\begin{align}\label{Teu_Det_Trans}
X=\frac{\sqrt{r^{2}+a^{2}}}{\Delta}\left[\alpha R^{(t)}+\frac{\beta}{\Delta}\frac{dR^{(t)}}{dr}\right] \,,
\end{align}
where the quantities $\alpha$ and $\beta$ are functions of $r$ and $\omega$, their explicit dependence can be found in~\cite{Detweiler:1977,Maggio:2018ivz}. With this transformation, it turns out that the Detweiler function satisfies the following Schr\"{o}dinger-like equation,
\begin{align}
\frac{d^{2}X}{dr_{*}^{2}}-V(r,\omega)X=0~,
\end{align}
where the potential is purely real and has the following asymptotics: at the horizon $V(r \to r_+, \omega) \to -\bar{\omega}^2$, and at infinity $V(r \to \infty, \omega) \to -\omega^2$. Hence, it is possible to define purely outgoing and ingoing wave modes $\sim e^{\pm i\bar{\omega} r_{*}}$ at the horizon and $\sim e^{\pm i\omega r_{*}}$ at infinity. Using these the Detweiler function near the boundary of the compact object reads,
\begin{equation}
X \sim e^{-i\bar{\omega}(r_{*}-r^{0}_{*})}+\mathcal{R}(\omega)e^{i\bar{\omega}(r_{*}-r_{*}^{0})}~, \quad r_* \sim r_*^0 \,, 
\end{equation}
where $r_{*}^{0}$ is the location of the surface of the object in the tortoise coordinate. Thus, we can define the reflectivity $\mathcal{R}(\omega)$ of the compact object in a straightforward manner as
\begin{align}
\mathcal{R}(\omega)=\left[\frac{1-\frac{i}{\bar{\omega}}\left(\frac{1}{X}\frac{dX}{dr_{*}}\right)}{1+\frac{i}{\bar{\omega}}\left(\frac{1}{X}\frac{dX}{dr_{*}}\right)} \right]_{r_{*}^{0}}~,
\end{align}
which is in terms of the Detweiler function and its first radial derivative near the surface of the compact object. To connect the ratio $(\mathcal{B}/\mathcal{A})$, appearing in~\ref{tidal_love_small}, to the reflectivity derived from the Detweiler function we first have to express the radial Teukolsky function in the $(v,r,\theta,\tilde{\phi})$ coordinates, in terms of the Boyer-Lindquist coordinates $(t,r,\theta,\phi)$. This has been performed explicitly in \ref{TeuDet}. Besides, one can use the transformation between the Teukolsky function and the Detweiler function, in order to determine the connection between the Detweiler reflectivity $\mathcal{R}$, with the Teukolsky reflectivity $(\mathcal{B}/\mathcal{A})$ (for a detailed expression, see \ref{TeuDet}). Finally, the above relation can be inverted, and the ratio $\mathcal{B}/\mathcal{A}$ can be expressed in terms of the Detweiler reflectivity $\mathcal{R}$, which when substituted back into~\ref{tidal_love_small}, will provide the TLNs in terms of the Detweiler reflectivity. In what follows we will consider some generic form of the reflectivity, focusing on its small-frequency expansion, as in~\ref{reflectivity}, and shall derive the dependence of the TLNs on the model parameters and frequency. In particular, our interest will be on the scaling of the TLN with the parameter $\epsilon$, providing the departure of the surface of the compact object from the BH horizon.  

\section{TLNs of compact objects}\label{sec:results}
Having set up the basic equations and boundary conditions, we 
compute the TLNs for (generically rotating) compact objects, both in the static and frequency-dependent case. 
As we shall show, our general framework provides results for static perturbations and non-spinning objects, which does \emph{not} coincide with the strictly static limit, more details will be discussed in Section \ref{static_TLN}. Here we will elucidate some general properties of the dynamical TLNs for rotating and non-rotating compact objects.

\subsection{The BH limit}

To set the stage, we start from the BH limit of our results, which corresponds to ${\cal R}=0$ and $\epsilon\to0$. We discuss the TLNs of BHs in the linear-frequency approximation and  comment on the behavior of the TLNs when terms of $\mathcal{O}(\omega^{2})$ are included in the analysis. 

At linear order in the frequency, one can directly use the analytical results derived in Sec.~\ref{sec:perturbations}. It is evident from~\ref{tidal_love_small} that the tidal response function of a rotating BH reads 
\begin{align}
F^{\rm rot, BH}_{\ell m}&=-i\left(\frac{am-2r_{+}M\omega}{r_+ - r_-}\right)\frac{(\ell-2)!(\ell+2)!}{(2 \ell)!(2 \ell+1)!} 
\nonumber
\\
&\prod_{j=1}^{\ell}\left[j^2+4\left(\frac{a m - 2 r_{+} M \omega}{r_+ - r_-}\right)^2 \right] \,,
\end{align}
which agrees with previous results~\cite{Chia:2020yla, Consoli:2022eey} (see also~\cite{Bhatt:2023zsy}). For a Schwarzschild BH, the tidal response function becomes    
\begin{equation}\label{bhnonrot}
F^{\rm nonrot,BH}_{\ell}=2iM\omega\frac{(\ell-2)!(\ell+2)!}{(2 \ell)!(2 \ell+1)!} 
\prod_{j=1}^{\ell}\left[j^{2}+16M^{2}\omega^{2}\right]\,.
\end{equation}
Note that in the non-rotating case, the tidal response function does not depend on the azimuthal number $m$, but only on the angular number $\ell$. For $\ell=2$, the quadrupolar response function for the Schwarzschild BH takes the form:
\begin{equation}
F^{\rm nonrot,BH}_{2}=\frac{i}{15}M\omega+{\cal O}(\omega^2M^2) \,.
\end{equation}
As evident in all of these cases, the response function in the small-frequency regime is a purely imaginary quantity and hence the TLNs, which are the conservative part of the response function and hence are given by the real part of the response function, identically vanishes. Our analysis is consistent with the findings of Ref.~\cite{Chia:2020yla} and shows that for BHs the TLNs identically vanish in the small frequency limit. 

At this outset we would like to emphasize that the above results are consistent with all the previous literature in the context of static TLNs. As evident from \ref{bhnonrot}, the TLNs of a non-rotating BH in GR identically vanishes up to linear order in $\omega$. This is consistent with our previous discussion, since for a BH there is no mismatch between the $\omega \to 0$ limit and the strictly static case. As we will demonstrate, this holds true for rotating BHs as well.

In the above expressions we have kept terms up to linear order in $\omega M$ and hence it is interesting to address what happens to the TLNs of BHs when terms $\mathcal{O}(\omega^{2}M^{2})$ are taken into account. The starting point is the radial Teukolsky equation where we keep all the terms up to $\mathcal{O}(\omega^{2}M^{2})$ in the near-horizon approximation (for a derivation, see \ref{quad_freq}), 
\begin{widetext}
\begin{align}
\frac{d^{2}R}{dz^{2}}&+\left[\frac{2iP_{+}-1}{z}-\frac{1+2iP_{+}+2i\omega[2M+(r_{+}-r_{-})]}{(1+z)}\right]\frac{dR}{dz}+\Bigg[-\frac{4iP_{+}}{z^{2}}+\frac{4iP_{+}+2i\omega[4M-(r_{+}-r_{-})]}{(1+z)^{2}}
\nonumber
\\
&\qquad -\frac{\ell(\ell+1)-2}{z(1+z)}+\frac{2ma\omega}{z(1+z)}\left(1-\frac{E_{1}}{2m}\right)
-\frac{2i\omega r_{+}}{z(1+z)}-\frac{\omega^{2}a^{2}(1+E_{2})}{z(1+z)}\Bigg]R=0~,
\end{align}
\end{widetext}
where the rescaled radial coordinate $z$ is defined in~\ref{defz}, and $E_1$ and $E_2$ are the coefficients of $\mathcal{O}(a\omega)$ and $\mathcal{O}(a^{2}\omega^{2})$ terms in the expansion of the angular eigenvalue in the angular Teukolsky equation, whose explicit forms can be obtained from~\cite{Berti:2005gp}. The first derivative term, namely $dR/dz$, does not depend on $\omega^{2}$, while the coefficient of the radial perturbation $R$ does only through the combination $a^{2}\omega^{2}$. {Hence it follows that for a Schwarzschild BH, any correction of $\mathcal{O}(\omega^{2}M^2)$ is absent in the Teukolsky equation. Still, it is possible that, due to non-analytical terms, the response function in the small-frequency limit, upon expanded up to the next-to-leading order, would contain terms $\mathcal{O}(\omega^{2}M^2)$, which will give rise to non-trivial TLNs of a Schwarzschild BH at quadratic order in the frequency. Such an analysis is currently underway through both analytical and numerical techniques, which will be reported elsewhere. Thus, the $\ell=2$ TLN of a Schwarzschild BH vanishes at least as
\begin{align}
k_{2}^{\textrm{nonrot, BH}}=0+\mathcal{O}(M^{2}\omega^{2})~.
\end{align}
The same holds for slowly rotating BHs to linear order in the spin since any possible correction to the TLNs is proportional to $\mathcal{O}(a\omega)$. 
These results agree with those of~\cite{Charalambous:2021mea, Bonelli:2021uvf, Chakrabarti:2013lua}.}
For a generically rotating BH, there can be non-trivial effects due to the $\mathcal{O}(a^{2}\omega^{2})$ and $\mathcal{O}(a\omega)$ term in the Teukolsky equation. In other words, while it is guaranteed that a BH has vanishing static TLNs, it could have non-trivial dynamical TLNs. We wish to explore this interesting problem in a future work. 

\subsection{Non-spinning case}

We present here the TLN of a non-rotating ultracompact object, which can be obtained by taking the limit $a\to 0$ in~\ref{tidal_love_small}. The resulting expression depends on the frequency, the mass of the compact object, the parameter $\epsilon$, and the reflectivity. In this limit $P_{+}=-2M\omega$, thus the expression for the TLN associated with the $\ell=2$ mode simplifies to, 
\begin{align}\label{tidal_love_nonrot}
k_{2}&=\textrm{Re}\Bigg[\left(\frac{iM\omega}{30}\right)\left(1+16M^{2}\omega^{2}\right)\left(1+4M^{2}\omega^{2}\right)
\nonumber
\\
&\qquad\qquad \times\left\{\frac{1-\frac{\mathcal{B}}{\mathcal{A}}\Gamma_{1}}{1+\frac{\mathcal{B}}{\mathcal{A}}\Gamma_{1}}\right\}\Bigg]~,
\end{align}
where (see~\ref{app:smallfrequency} for a derivation)
\begin{align}\label{Gamma1_freq}
\Gamma_{1}
=\frac{3i(1-2iM\omega)}{M\omega(1+4M^{2}\omega^{2})(1+16M^{2}\omega^{2})}~.
\end{align}
As we will demonstrate later, at frequencies satisfying the condition $|1+(\mathcal{B}/\mathcal{A})\Gamma_{1}|\simeq 0$, there are resonances in the TLNs. To see this explicitly, one must use the relation between the Teukolsky reflectivity $(\mathcal{B}/\mathcal{A})$ with the Detweiler reflectivity $\mathcal{R}(\omega)$, which in the non-rotating case reads (see \ref{app:smallfrequency} for details), 
\begin{align}\label{Teu_Det_Rel}
\frac{\mathcal{B}}{\mathcal{A}}&=\left(\frac{2M\omega}{3}\right)\frac{(i+2M\omega+16iM^{2}\omega^{2}+32M^{3}\omega^{3})}{2-iM\omega}
\nonumber
\\
&\qquad \times \mathcal{R(\omega)}e^{8\pi M\omega-2i\omega\left(r_{*}^{0}-2M\right)}~.
\end{align}
In the next sections, we will discuss the zero-frequency limit of the above expressions and hence determine the static TLNs from the dynamical ones. 

\subsubsection{Zero-frequency limit}
In this section, we discuss the zero-frequency limit of the TLNs of static and spherically symmetric compact objects.
Before proceeding, it is worth emphasizing that we need to set the rotation parameter $a\to 0$ and then the frequency $\omega \to 0$ to determine the static TLN for a non-rotating compact object. As we will show in detail, the zero-frequency limit of the frequency-dependent TLN differs from the strictly static TLN, as the branch of the solution to the Teukolsky equation with outgoing behavior close to the horizon becomes ill-defined in this limit. Therefore, we do not expect the TLN in the zero frequency limit to match with the results derived in Ref.~\cite{Cardoso:2017cfl}, but nonetheless the zero-frequency limit seems more natural to consider as the coalescence of binary compact objects is frequency-dependent.

Given the expression for $\Gamma_{1}$ in~\ref{Gamma1_freq} and the relation between the Teukolsky and Detweiler reflectivities in~\ref{Teu_Det_Rel}, we can easily obtain the zero-frequency limit of both of these expressions, by keeping the leading order terms in $M\omega$. This yields $\Gamma_{1}=(3i/M\omega)+6+\mathcal{O}(M\omega)$, while the Teukolsky reflectivity reduces to
\begin{align}
\frac{\mathcal{B}}{\mathcal{A}}&=\frac{iM\omega\mathcal{R}}{3}+\frac{M^2 \omega^2 \mathcal{R}}{6}\left(3 + 16 i \pi + 8\epsilon + 8 \ln \epsilon\right)
\nonumber
\\
&+\mathcal{O}(M^{3}\omega^{3})~.
\end{align}
Thus, keeping terms up to linear order in $M\omega$, we obtain 
\begin{align}
\left(\frac{\mathcal{B}}{\mathcal{A}}\right)\Gamma_{1}=-\mathcal{R}+\frac{i M \omega \mathcal{R}}{2} \left( 7 + 16 i \pi + 8\epsilon + 8\ln \epsilon\right)\,,
\end{align}
and hence the TLN in the zero frequency limit becomes,
\begin{align}\label{tidal_love_nonrot2}
k_{2} &=\textrm{Re}\Bigg\{\left(\frac{iM\omega}{30}\right) 
\nonumber
\\
&\times\frac{1+\mathcal{R(\omega)}\left[1-\frac{iM\omega}{2}\left(7+16i\pi+8\epsilon+8\ln \epsilon\right)\right]}{1-\mathcal{R}(\omega)\left[1-\frac{iM\omega}{2}\left(7+16i\pi+8\epsilon+8\ln \epsilon\right)\right]}\Bigg\}~.
\end{align}
The above equation shows one of our main results: except for the single case where the above expression has a pole (to be discussed below), the TLN identically vanishes in the zero-frequency limit for any other reflectivity, just as in the BH case. This includes all cases of partial reflection\footnote{It might come as a surprise that the TLNs are zero if ${\cal R}\neq0$ in the zero-frequency limit. For example one could argue that a perfect-fluid star has nonzero TLNs and viscosity (which effectively corresponds to $|{\cal R(\omega)}|^2<1$) cannot drastically change the TLNs. However, note that viscosity would affect the reflectivity starting at ${\cal O}(M\omega)$, leaving ${\cal R}_0$ unaffected. Thus, our formalism also accounts for the TLNs of a neutron star, wherein ${\cal R}_1$ depends on the equation of state.}, i.e., $|{\cal R}(\omega)|^2<1$. In the exceptional case where $\mathcal{R}(\omega)=1+iM\omega \mathcal{R}_{1}$,~\ref{tidal_love_nonrot2} has a pole and the TLN for the $\ell=2$ mode, in the zero-frequency limit, becomes
\begin{align}\label{tidal_love_nonrot_f}
k_2=\frac{2}{15} \textrm{Re}\Bigg[\frac{1}{-2\mathcal{R}_{1}+\left\{7 + 16 i \pi + 8 (\epsilon + \ln \epsilon)\right\}}\Bigg]~.
\end{align}
The above result does not correspond to the strictly static case, as we must keep terms $\mathcal{O}(M\omega)$ in the combination $(\mathcal{B}/\mathcal{A})\Gamma_{1}$ as well as the reflectivity $\mathcal{R}$ must be frequency-dependent (i.e., $\mathcal{R}_1 \neq 0$) in order to have a non-zero TLN. 
Note that, the magnitude of the reflectivity reads $|\mathcal{R}(\omega)|=1-M\omega\, \textrm{Im}\mathcal{R}_{1}+\mathcal{O}(\omega^{2})$. Thus, if $\mathcal{R}_{1}$ is real, the object is perfectly reflecting, i.e., $|\mathcal{R}(\omega)| =1$. In general, if $\mathcal{R}_{1}$ is complex with $\textrm{Im}\mathcal{R}_{1}>0$, the reflectivity is smaller than unity. Note that $\textrm{Im}\mathcal{R}_{1}<0$ is not physical since it corresponds to a reflectivity larger than unity. Thus, non-zero TLNs of compact objects require the zero-frequency reflectivity to satisfy $\mathcal{R}_{0}=1$, which we expect to hold generically for non-dissipative compact objects including perfect-fluid stars.
This expectation is supported by previous work involving perfectly reflecting compact objects (see e.g., \cite{Cardoso:2017cfl}), for which $\mathcal{R}_{0}=1$ and $\mathcal{R}_{1}=0$, yielding non-zero TLNs. 
Furthermore, this result is expected on physical grounds: a non-dissipative object should have perfect reflectivity for a zero-frequency wave, with no phase shift, hence $\mathcal{R}_{0}=1$.\footnote{In a separate work~\cite{MichelaTesi} we apply this formalism to specific examples in the non-spinning case, including perfect-fluid neutron stars and exotic compact objects, explicitly computing the complex function $\mathcal{R}(\omega)$. It turns out that $\mathcal{R}_{0}=1$ in all of these models with non-zero TLNs, and that $\mathcal{R}_{1}$ is precisely the value yielding the expected TLNs.}

It is important to stress that, the TLN associated with the $\ell=2$ mode depends on the frequency-dependent part of the reflectivity and in general displays a logarithmic behavior with $\epsilon$. Whenever $|\ln\epsilon|\gg |{\cal R}_1|/4$, in the $\epsilon\to0$ limit, the TLN vanishes as
\begin{equation}
k_2=\frac{1}{60\ln\epsilon}\,.
\end{equation}
Although in this case we obtain a logarithmic scaling as found in~\cite{Cardoso:2017cfl} for static TLNs of perfectly reflecting ECOs, the numerical coefficients differ from those obtained in the zero-frequency limit of dynamical TLNs.
Thus, the static limit of the dynamical TLN does not coincide with the strictly static case. This is because the solution of the Teukolsky equation in the zero-frequency case differs from the one with linear-in-frequency terms. We will discuss these aspects in detail in Sec.~\ref{static_TLN}.
{The above logarithmic scaling of the TLNs has also appeared in the effective field theory computations involving tidal effects \cite{Saketh:2023bul, Mandal:2023hqa}. Though the origin of these terms in \cite{Saketh:2023bul, Mandal:2023hqa} are associated with renormalization in effective field theories, while here the logarithmic behaviour arises from the reflective nature of the compact object. It will be worth pursuing to understand a connection between the two, in particular, if $\epsilon$ provides a natural cut-off scale for the effective field theory computation, leading to the characteristic logarithmic behaviour.}

\begin{figure}[!t] 
\centering
\includegraphics[width=\columnwidth]{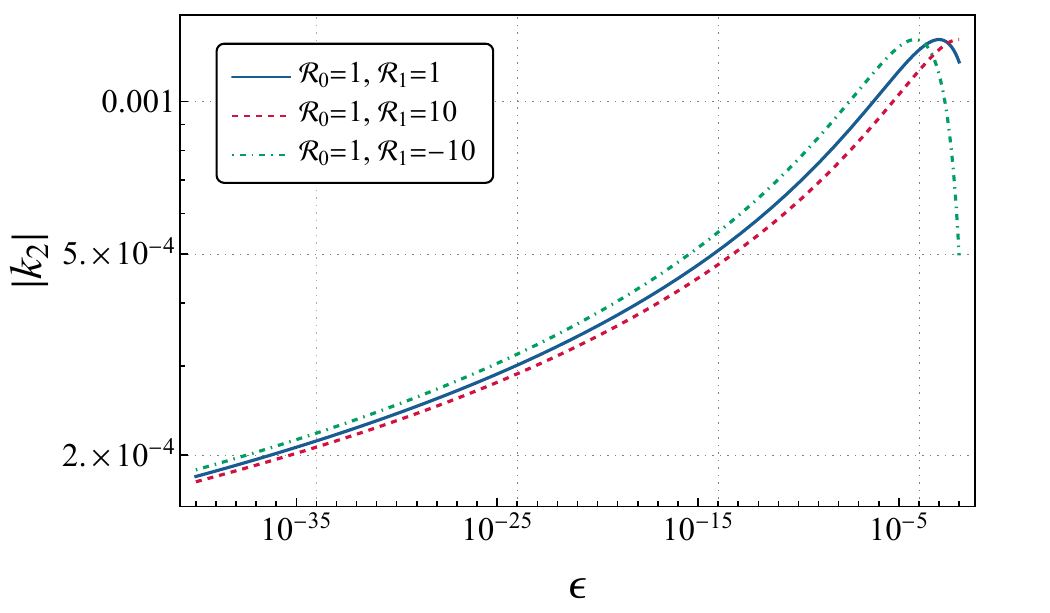}
\caption{Absolute value of the quadrupolar TLN, $|k_{2}|$,  the zero-frequency limit and as a function of the compactness parameter $\epsilon$, in a logarithmic scale, and for $\mathcal{R}_{0}=1$ and  various choices of the linear-in-frequency reflectivity $\mathcal{R}_{1}$. For larger and positive values of $\mathcal{R}_{1}$ the absolute value of the TLN decreases, while for negative choices of $\mathcal{R}_{1}$ the TLN increases. Overall, as $\epsilon$ becomes larger, the TLN also increases. The plot shows the generic logarithmic dependence of the static TLN on the compactness parameter $\epsilon$.}
\label{fig:a0_omega0_perfect}
\end{figure}

The variation of the dynamical TLN in the zero-frequency limit with $\epsilon$ is presented in \ref{fig:a0_omega0_perfect}, for $\mathcal{R}_{0}=1$, and three different choices of $\mathcal{R}_{1}$. The plot clearly demonstrates the dependence of the static TLN on $\log \epsilon$, for constant values of $\mathcal{R}_{1}$. Also, \ref{fig:a0_omega0_perfect} shows that the magnitude of the static TLN depends heavily on $\epsilon$ and also mildly on $\mathcal{R}_{1}$, as suggested by \ref{tidal_love_nonrot_f}.

\subsubsection{Frequency dependence of the TLNs}

\begin{figure}[!t] 
\centering
\includegraphics[width=\columnwidth]{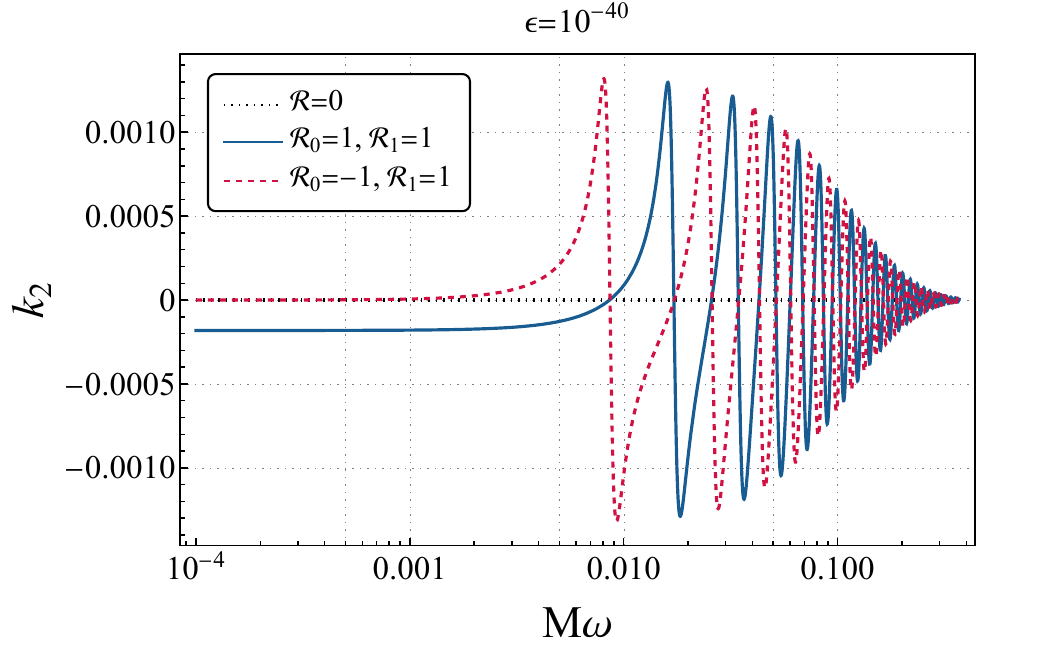}
\caption{Quadrupolar TLN as a function of the frequency, for three cases --- (a) BHs, with $\mathcal{R}=0$, (b) for a compact object, with $\mathcal{R}_{0}=1$ and (c) another compact object with $\mathcal{R}_{0}=-1$. In the latter two cases, involving non-rotating compact objects we have taken $\epsilon=10^{-40}$ and $\mathcal{R}_{1}=1$. As evident, only for the choice $\mathcal{R}_{0}=1$, the TLN goes to a finite and negative value in the zero-frequency limit, while even for $\mathcal{R}_{0}=-1$ (which still corresponds to a perfectly reflecting object), the TLN vanishes in the zero-frequency limit. In both of these cases, the TLN increases with an increase in frequency, and in general, the frequency-dependent TLN does not vanish.}
\label{fig:a0_omega0_perfect2}
\end{figure} 

\begin{figure}[!t] 
\centering
\includegraphics[width=\columnwidth]{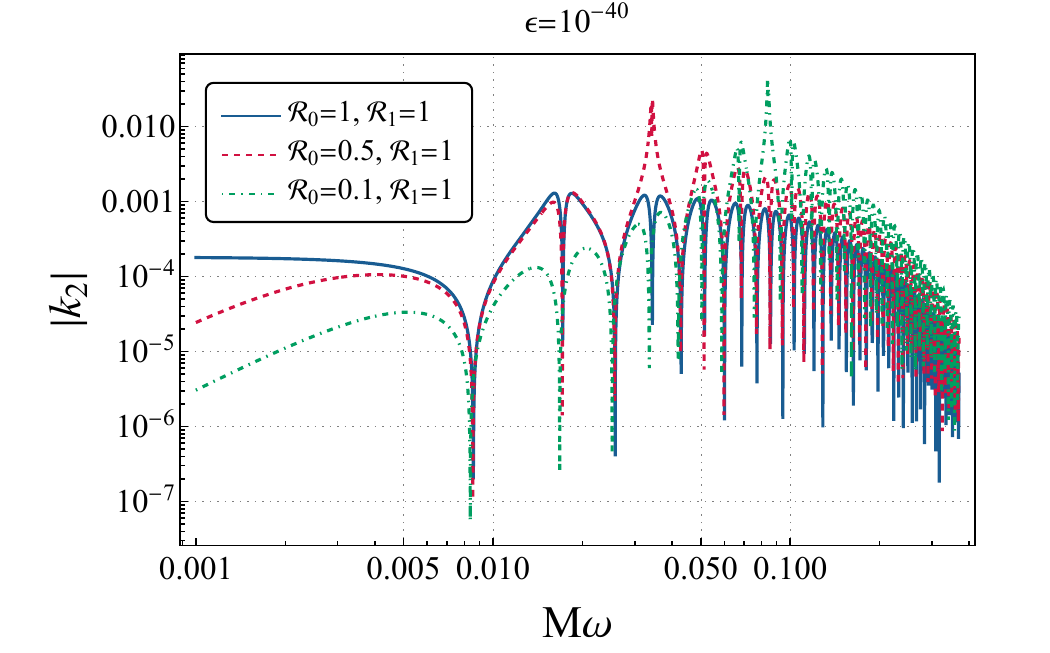}
\caption{The TLN of a compact object as a function of the frequency for a non-spinning and partially as well as perfectly reflecting compact object with $\epsilon=10^{-40}$ and $\ell=2$. As evident, except for the case $\mathcal{R}_{0}=1$, the TLN in the zero-frequency limit identically vanishes. On the other hand, for $\mathcal{R}_{0}=1$, there are oscillations at non-zero frequencies, while for $\mathcal{R}_{0}<1$, there are  also resonances, which drift to higher frequencies as the reflectivity decreases. See the main text for more discussion.}
\label{fig:a0_omega0_partial}
\end{figure}

Having discussed the static limit of the dynamical TLNs for non-rotating compact objects, in this section, we will discuss the properties of the dynamical TLN associated with the $\ell=2$ mode. By combining~\ref{Gamma1_freq} and~\ref{Teu_Det_Rel}, the expression for the TLN associated with the $\ell=2$ mode of a non-rotating compact object reads, 
\begin{align}\label{nonrotfreq}
k_{2}&=\textrm{Re}\Bigg[\left(\frac{iM\omega}{30}\right)\left(1+16M^{2}\omega^{2}\right)\left(1+4M^{2}\omega^{2}\right)
\nonumber
\\
&\qquad\qquad \qquad \times \left(\frac{1+\mathcal{R}(\omega)G(\omega)}{1-\mathcal{R}(\omega)G(\omega)}\right)\Bigg]~,
\end{align}
where we have introduced the frequency-dependent quantity $G(\omega)$,
\begin{align}
G(\omega)&\equiv 2\frac{\exp\left[8\pi M\omega-4iM\omega\left(\epsilon+\ln \epsilon \right)\right]}{2-iM\omega}
\nonumber
\\
&\qquad \times \left(\frac{1-2iM\omega+16M^{2}\omega^{2}-32iM^{3}\omega^{3}}{1+2iM\omega+16M^{2}\omega^{2}+32iM^{3}\omega^{3}}\right)~.
\end{align}
Note that $G(\omega)\to 1$ in the zero-frequency limit, so one recovers~\ref{tidal_love_nonrot2}.
The frequency dependence of the TLN for a non-rotating compact object, both in the perfectly reflecting and partially absorbing cases have been depicted in \ref{fig:a0_omega0_perfect2} and~\ref{fig:a0_omega0_partial}, respectively. As evident, except for the case $\mathcal{R}_0=1$, the TLN identically vanishes in the limit $\omega\rightarrow 0$, which is consistent with~\ref{tidal_love_nonrot2}. Moreover, the TLN associated with the $\ell=2$ mode in the zero-frequency limit is negative for $\mathcal{R}_{0}=1$ and a positive choice of $\mathcal{R}_{1}$ and small $\epsilon$, which is consistent with~\ref{tidal_love_nonrot2}. We will now discuss the features associated with the frequency-dependent TLNs.

In all the cases involving either a perfectly reflecting compact object with $\mathcal{R}_{0}=\pm 1$ (see \ref{fig:a0_omega0_perfect2}), or a partially reflecting one with $\mathcal{R}_{0}<1$ (see \ref{fig:a0_omega0_partial}), the TLN associated with the $\ell=2$ mode increases with frequency, and we observe oscillations. \ref{fig:a0_omega0_partial} also suggests that the $\ell=2$ TLN decreases at higher frequencies, but our low-frequency analysis, presented here, eventually breaks down, and thus the high-frequency results should be taken with a grain of salt. 

We will now depict the one-to-one correspondence between the oscillation frequencies as well as the resonances with the real part of the QNMs of the compact object. Following Ref.~\cite{Maggio:2018ivz}, the QNMs in the small-$\epsilon$ limit can be written as,
\begin{equation}\label{QNManalytics}
\omega_R \sim -\frac{\pi (q+1)}{2 |r_*^0| } + m \Omega_{\rm H} \,,
\end{equation}
where $r_*^0 \sim M [1+(1-\chi^2)^{-1/2}] \ln\epsilon$, $\chi=a/M$ is the dimensionless spin,   $q=2n-1 \ (2n)$ for perfectly reflecting objects with $\mathcal{R}_0=1$ ($\mathcal{R}_0=-1$), and $n \geq 1$ is the overtone number. As discussed in Refs.~\cite{Maggio:2018ivz, Maggio:2017ivp,Maggio:2021uge}, the above analytical approximation for the real part of the QNMs is also valid for partially absorbing  objects, with $q=2n-1 \ (2n)$ for compact objects with $\mathcal{R}_0>0$ ($\mathcal{R}_0<0$).
In the case of a non-spinning compact object with $\mathcal{R}_0>0$ and $\epsilon=10^{-40}$, the analytical approximation for the real part of the QNMs in~\ref{QNManalytics} yields the following frequencies $M\omega_R = 0.017 \ (n=1), 0.034 \ (n=2), 0.051 \ (n=3), 0.068 \ (n=4), 0.085 \ (n=5)$, etc. Intriguingly, at precisely these frequencies the oscillations in the dynamical TLN occur, as shown  in~\ref{fig:a0_omega0_partial}. The amplitudes of the oscillations and resonances are then related to the absolute value of $\mathcal{R}_0$ and the factor $G(\omega)$.
As evident from \ref{nonrotfreq}, for partially reflective compact objects, close to the point $\mathcal{R}(\omega)G(\omega)\simeq 1$, the TLN for the $\ell=2$ mode displays resonances. This can be verified from~\ref{fig:a0_omega0_partial} as well, which shows the appearance of resonances at two different frequencies for the choices of reflectivities considered here. One can easily verify that the locations of these resonances are exactly at the solutions of the equation $\mathcal{R}(\omega)G(\omega)\simeq 1$, as predicted by the analytical results. In particular, for smaller values of the reflectivity, these resonances appear at larger values of $M\omega$, and hence, as the reflectivity becomes very small, the resonance frequencies will become so large that the small-frequency approximation will break down. We also observe that the amplitudes of the resonances become larger for smaller reflectivities, since the overall multiplication factor involving $M\omega$ in the TLN becomes larger.

The above summarizes the behavior of the TLNs for non-rotating compact objects both in the dynamical context, as well as in the zero-frequency limit. We now turn our attention to the case of spinning compact objects.    
\subsection{Spinning case}

Having discussed the features of the TLN in the case of a nonspinning compact object, in this section, we will concentrate on determining the TLN in the spinning case. The general expression has been provided in~\ref{tidal_love_small}, but we will quote some results associated with the $\ell=2$ mode since that is the most relevant one. For this case, the tidal response function can be determined using \ref{overall_factor},~\ref{Gamma_1} and~\ref{Gamma_2} in~\ref{app:smallfrequency}, which reads, 
\begin{align}\label{tln_rot_freq}
k_{2m}&=\textrm{Re}\Bigg[-\frac{iP_{+}}{60}\left(1+4P_{+}^{2}\right)\left(1+P_{+}^{2}\right)
\nonumber
\\
&\qquad \qquad \times \left\{\frac{1+\left(\frac{\mathcal{B}}{\mathcal{A}}\right)\frac{6i\left(1+iP_{+}\right)}{P_{+}\left(1+4P_{+}^{2}\right)\left(1+P_{+}^{2}\right)}}{1-\left(\frac{\mathcal{B}}{\mathcal{A}}\right)\frac{6i\left(1+iP_{+}\right)}{P_{+}\left(1+4P_{+}^{2}\right)\left(1+P_{+}^{2}\right)}}\right\}\Bigg]~,
\end{align}
where $P_{+}=\{(am-2M\omega r_{+})/(r_{+}-r_{-})\}$ as defined in \ref{Pplusminus}. Note that, in the non-rotating case, it was possible to derive a relation between the Teukolsky reflectivity $(\mathcal{B}/\mathcal{A})$ and the Detweiler reflectivity $\mathcal{R}$ in a closed form, see \ref{Teu_Det_Rel}. However, in the rotating case, the analytical expression for such a relation is cumbersome and we will not report it. Following this procedure, we express the TLN in~\ref{tln_rot_freq} in terms of the more physical Detweiler reflectivity. Note that, unlike the non-rotating case, the TLN for a rotating compact object depends on the azimuthal number $m$ through $P_{+}$. Moreover, even when the $\omega\rightarrow 0$ limit is taken, due to the non-zero rotation parameter, $P_{+}$ remains non-zero and hence the TLN may not vanish. However, the exact zero-frequency limit depends on the relation between the Teukolsky and the Detweiler function in the appropriate limits. In what follows, we will depict the dependence of the TLN with the frequency as well as with $\epsilon$, capturing the departure of the surface of the compact object from the would-be horizon, for various choices of the reflectivity and rotation parameter. 

\begin{figure}[!t] 
\centering
\includegraphics[width=\columnwidth]{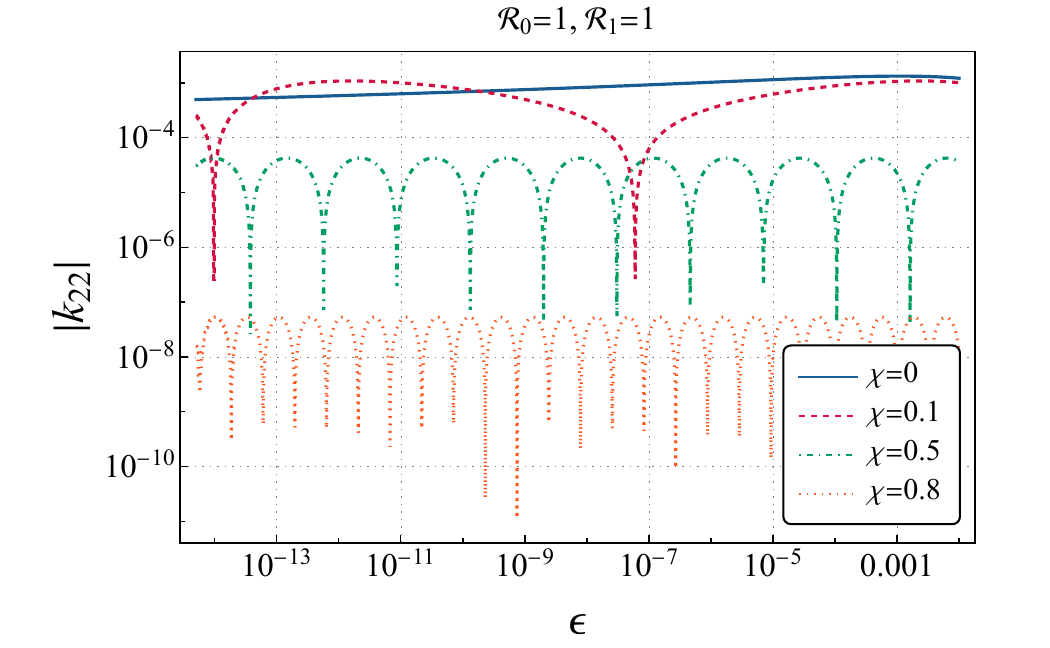}
\caption{The zero-frequency limit of the frequency-dependent TLN as a function of $\epsilon$, which determines the distance between the surface of the compact object and the would-be horizon for a spinning and perfectly reflecting compact object ($\mathcal{R}_{0}=1=\mathcal{R}_{1}$) with $\ell=2$ and $m=2$. We  considered three possible choices of the dimensionless rotation parameter $\chi$. It is clear that, increasing the rotation,  the zero-frequency TLN decreases.}
\label{fig:an0_epsilon}
\end{figure}

\begin{figure}[!t] 
\centering
\includegraphics[width=\columnwidth]{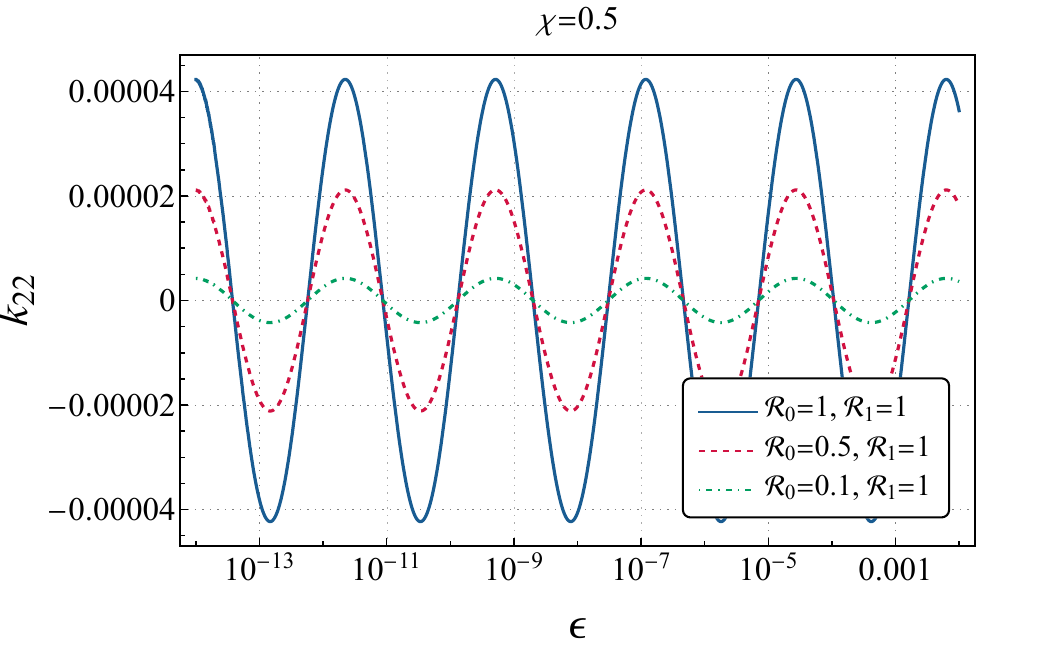}
\caption{Zero-frequency limit of the TLN as a function of $\epsilon$ for a spinning compact object with $\chi=0.5$, $\ell=2$, $m=2$ and for three different choices of the reflectivity. The plot shows that a decrease in the zero-frequency reflectivity $\mathcal{R}_{0}$ decreases the TLN, while the oscillation pattern remains the same.}
\label{fig:an0_reflective}
\end{figure}

The dependence of the TLN on $\epsilon$, the parameter which determines the distance between the reflective surface and the would-be horizon, for a rotating compact object, at zero-frequency is presented in~\ref{fig:an0_epsilon} and~\ref{fig:an0_reflective}, respectively. In~\ref{fig:an0_epsilon}, we have plotted the absolute value of the TLN against $\epsilon$ for a perfectly reflecting compact object, with $\mathcal{R}_{0}=1=\mathcal{R}_{1}$, and for various choices of the dimensionless rotation parameter. Interestingly, the zero-frequency TLN decreases with increasing rotation, i.e., faster-rotating compact objects are more difficult to deform than the slowly rotating ones. Furthermore, the TLN decreases as the reflectivity of the compact object decreases, as expected (cf. \ref{fig:an0_reflective}). This is because, the $\mathcal{R}_{0}\rightarrow 0$ limit corresponds to the Kerr BH, having zero TLN at least for small frequencies, and hence it is expected that a smaller zero-frequency reflectivity leads to a smaller TLN. We would like to point out that the oscillatory behavior of the TLN with $\epsilon$ originates from the term involving $\exp[-2i\omega(r_{0}^{*}-r_{+})]$, appearing in the connection between the Teukolsky reflectivity $(\mathcal{B}/\mathcal{A})$ and the Detweiler reflectivity $\mathcal{R}$. Further, as evident from \ref{fig:an0_reflective}, the oscillation length scale does not depend on the reflectivity, i.e., all the curves in the figure are in phase with each other. However, the amplitude depends on the reflectivity and decreases for smaller reflectivity. Note that both of these plots are in the zero-frequency limit, and hence it is clear that, unlike the case of a non-rotating ECO where the TLN generically vanishes in the zero-frequency limit, for a rotating ECO the zero-frequency TLN is in general non-zero. 

\begin{figure}[!t] 
\centering
\includegraphics[width=\columnwidth]{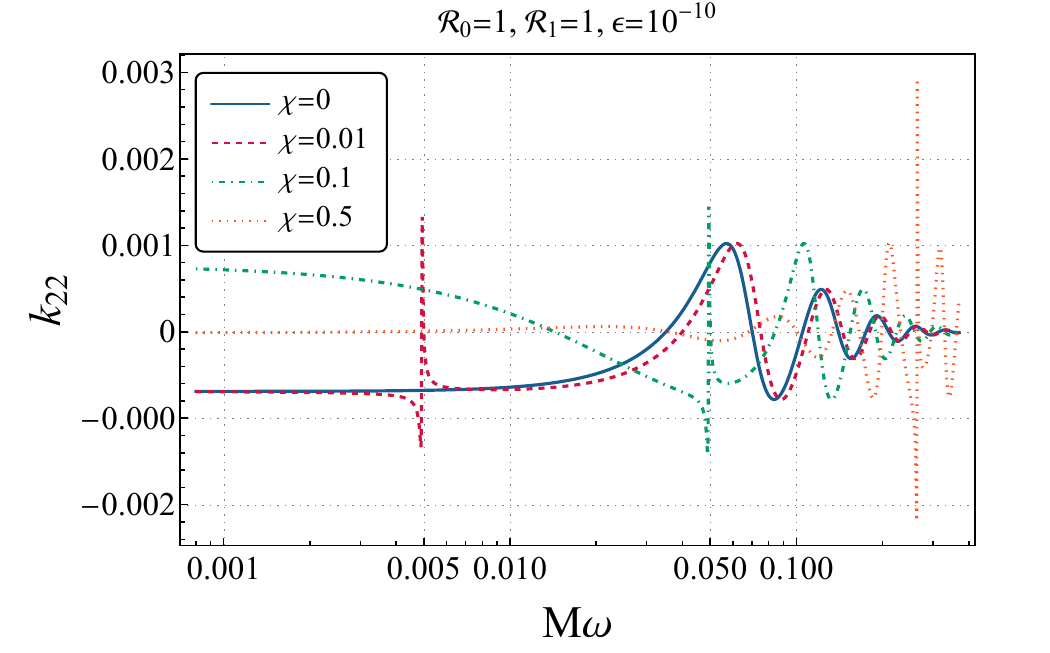}
\caption{TLN of a spinning and perfectly reflecting $(\mathcal{R}_{0}=1=\mathcal{R}_{1})$ compact object as a function of the frequency with $\epsilon=10^{-10}$, $\ell=2$ and $m=2$. We observe resonances in the spectrum of the TLN as in the case of non-rotating compact objects. Interestingly, with an increase in rotation, the TLN decreases significantly, validating our claim that faster-rotating objects are difficult to deform tidally at lower frequencies.}
\label{fig:an0_omegan0_spin}
\end{figure}

\begin{figure}[!t] 
\centering
\includegraphics[width=\columnwidth]{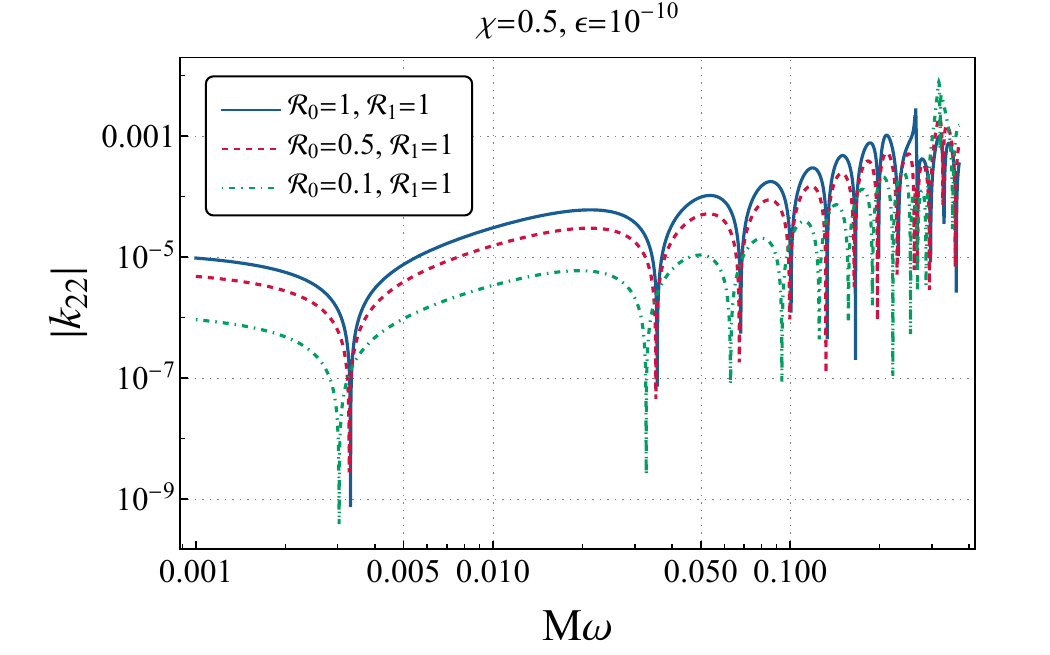}
\caption{The absolute value of the TLN  as a function of the frequency for spinning compact objects with three different choices of the reflectivity and the following choices of the physical parameters: $\chi=0.5$, $\epsilon=10^{-10}$, $\ell=2$ and $m=2$. Following our expectations, a decrease in the reflectivity results in a decrease in the TLN, ultimately vanishing in the BH limit.}
\label{fig:an0_omegan0_reflect}
\end{figure}

Besides the zero-frequency limit, we have also studied the dependence of the $\ell=2$ TLN on the frequency. In \ref{fig:an0_omegan0_spin}, we show this dependence for different choices of the spin, for a perfectly reflecting compact object with $\mathcal{R}_{0}=1=\mathcal{R}_{1}$ and $\epsilon=10^{-10}$. The plot explicitly demonstrates the presence of resonances in the structure of the TLN, and these precisely appear where the denominator of \ref{tln_rot_freq} vanishes. The resonance frequencies associated with the dynamical TLNs also correspond to the QNM frequencies of the rotating object, as discussedi n the previous section.  In the dynamical context, for higher frequencies, the behavior of the TLN is very similar to that of a non-rotating compact object, involving oscillations and resonances at specific frequencies. Thus, one can safely argue that the frequency-dependent TLN of a compact object has similar behavior in both the non-rotating and rotating case. In particular, rapidly rotating compact objects have larger TLNs at higher frequencies, but very small TLNs for low-frequency tidal fields. Therefore, at lower frequencies faster-rotating objects are difficult to be tidally deformed, while at slightly higher frequencies faster-rotating objects are the easiest to be tidally deformed. On a similar note, \ref{fig:an0_omegan0_reflect} depicts the variation of the $\ell=2$ TLN with the frequency for various partially reflecting rotating compact objects. In this case, the TLN also decreases with reflectivity, implying that objects with smaller reflectivity are difficult to deform. This result is in complete agreement with the fact that BHs, which are objects with no reflectivity, cannot be tidally deformed. Thus, we have provided all the details regarding the TLNs of a rotating compact object in the dynamical context, as well as in the zero frequency limit, depicting non-trivial behavior. 

\subsection{Static TLNs for a non-rotating compact object}\label{static_TLN}

As discussed in the previous sections, the strictly static TLN for a non-rotating compact object needs to be derived from scratch. One of the prime reasons is the fact that the outgoing solution of the Teukolsky equation derived in \ref{app:smallfrequency} is ill-behaved in the limit $P_{+}\to 0$ (as we are interested in non-rotating objects with $a=0$, and the static limit correspond to $\omega \to 0$). Thus, the static limit for a non-rotating compact object needs to be studied by taking the zero-rotation and zero-frequency limit of the Teukolsky equation, presented in \ref{Teukeq}, which reads, 
\begin{align}\label{teuk_static_n}
\dfrac{d^{2}R_{\ell}}{dr^{2}}&+\left(-\frac{1}{r-r_{+}}-\frac{1}{r-r_{-}}\right)\dfrac{dR_{\ell}}{dr}
\nonumber
\\
&+\left(-\frac{1}{r-r_{+}}+\frac{1}{r-r_{-}}\right)\frac{\gamma_{\ell}}{(r_{+}-r_{-})}R_{\ell}=0~,
\end{align}
where, we have defined $\gamma_{\ell}\equiv(\ell+2)(\ell-1)$. Introducing, the variable $z$ as in \ref{app:smallfrequency}, in the present scenario we obtain $z=\frac{r}{2M}-1$, 
in terms of which the above differential equation reduces to, 
\begin{align}
\dfrac{d^{2}R_{\ell}}{dz^{2}}-\left(\frac{1}{z}+\frac{1}{1+z}\right)\dfrac{dR_{\ell}}{dz}+\left(-\frac{1}{z}+\frac{1}{1+z}\right)\gamma_{\ell}R_{\ell}=0~.
\end{align}
This equation has the following explicit solution,
\begin{align}\label{staticTeu}
R_{\ell}=z(1+z)\left[c_{1}P_{\ell}^{2}(1+2z)+c_{2}Q_{\ell}^{2}(1+2z)\right]~,
\end{align}
where, $P_{\ell}^{2}(x)$ and $Q_{\ell}^{2}(x)$ are the associated Legendre polynomials. To see explicitly how these solutions differ from the zero-frequency and zero-rotation limit of \ref{gen_sol}, we use the relations between the associated Legendre polynomials and the hypergeometric functions,  
\begin{align}
P_{\ell}^{2}(1+2z)=\frac{\Gamma(\ell+3)}{2\Gamma(\ell-1)}z(1+z)\,_{2}F_{1}(3+\ell,2-\ell;3;-z)~. 
\end{align}
As evident, the above hypergeometric function  follows from the first solution in \ref{gen_sol}, under the limit $P_{+}\to 0$. While, for the other associated Legendre polynomial, 
\begin{align}
Q_{\ell}^{2}(1+2z)&=\frac{\sqrt{\pi}}{2^{\ell-1}}\frac{\Gamma(\ell+3)}{2\Gamma(\ell+\frac{3}{2})}\frac{z(1+z)}{(1+2z)^{\ell+3}}
\nonumber
\\
&\times\,_{2}F_{1}\left(2+\frac{\ell}{2},\frac{3}{2}+\frac{\ell}{2};\ell+\frac{3}{2};\frac{1}{(1+2z)^{2}}\right)~,
\end{align}
which has no analog in the static limit of \ref{gen_sol} for a non-rotating system. This is why the solution to the Teukolsky equation in the strictly static situation differs from the static limit of the dynamical solution for the perturbation of a reflecting compact object. 

The above solution for the radial part of the Teukolsky function $r^{4}\Psi_{4}$ holds for any values of $z$, and hence we can start by considering the near-horizon, i.e., $z\to 0$ limit, as well as the far zone, which corresponds to the large-$z$ limit of it. Let us first explore the far-zone region. In this case, the associated Legendre polynomial $P_{\ell}^{2}(1+2z)$ can be expanded as,
\begin{align}
P_{\ell}^{2}(1+2z)\Big|_{\rm far}\simeq \frac{\Gamma(\ell+3)}{2\Gamma(\ell-1)}z^{2}&\Big[\frac{\Gamma(3)\Gamma(-1-2\ell)}{\Gamma(2-\ell)\Gamma(-\ell)}z^{-3-\ell}
\nonumber
\\
&+\frac{\Gamma(3)\Gamma(1+2\ell)}{\Gamma(\ell+3)\Gamma(1+\ell)}z^{\ell-2}\Big]~.
\end{align}
Even though the ratio $\{\Gamma(-1-2\ell)/\Gamma(-\ell)\}$ has a finite limit for integer values of $\ell$, the term $\Gamma(2-\ell)$ diverges for any $\ell\geq 2$, and hence, the coefficient of the term $z^{-3-\ell}$ identically vanishes. Therefore, for $P_{\ell}^{2}$, we obtain the following result in the far-zone limit, 
\begin{align}
P_{\ell}^{2}(1+2z)\Big|_{\rm far}=\frac{\Gamma(1+2\ell)}{\Gamma(\ell-1)\Gamma(1+\ell)}\left(\frac{r}{2M}\right)^{\ell}~,
\end{align}
such that for $\ell=2$, we obtain, 
\begin{align}
P_{2}^{2}(1+2z)\Big|_{\rm far}=3\left(\frac{r}{M}\right)^{2}~,
\end{align}
which matches exactly with the expressions derived in Ref.~\cite{Hinderer:2007mb}. On a similar footing, the other associated Legendre polynomial $Q_{\ell}^{2}$ in the far-zone reads,
\begin{align}
Q_{\ell}^{2}(1+2z)\Big|_{\rm far}=\frac{\sqrt{\pi}\Gamma(3+\ell)}{2^{2\ell+2}\Gamma(\ell+\frac{3}{2})}z^{-\ell-1}~.
\end{align}
Thus, among the two independent solutions of the Teukolsky equation, $P_{\ell}^{2}$ provides the tidal field, while $Q_{\ell}^{2}$ yields the response of the non-rotating compact object to the tidal field. Note that, for $\ell=2$, the far-zone limit of $Q_{2}^{2}(1+2z)$ reduces to $(8/5)(M/r)^{3}$, which is again consistent with~\cite{Hinderer:2007mb}. Therefore, the radial part of the Teukolsky function, from \ref{staticTeu}, can be expressed as,
\begin{align}
R_{\ell}\Big|_{\rm far}&=c_{1}\frac{\Gamma(1+2\ell)}{\Gamma(\ell-1)\Gamma(1+\ell)}\left(\frac{r}{2M}\right)^{\ell+2}
\nonumber
\\
&\qquad \times \left[1+\mathcal{F}_{\ell}\left(\frac{r}{2M}\right)^{-2\ell-1}\right]~.
\end{align}
Here, $\mathcal{F}_{\ell}$ is the tidal response function, which is purely real, and thus the TLN becomes, 
\begin{align}
k_{\ell}=\frac{c_{2}}{c_{1}}\frac{\sqrt{\pi}\Gamma(3+\ell)}{2^{2\ell+3}\Gamma(\ell+\frac{3}{2})}\frac{\Gamma(\ell-1)\Gamma(1+\ell)}{\Gamma(1+2\ell)}~.
\end{align}
Therefore the determination of the TLN boils down to determining the ratio $(c_{2}/c_{1})$. For BHs, one can show that at the horizon, the associated Legendre polynomials $P_{\ell}^{2}$ read
\begin{align}
P_{\ell}^{2}(1+2z)\Big|_{\rm near}&=\frac{\Gamma(\ell+3)}{2\Gamma(\ell-1)}z(1+z)
\nonumber
\\
&=\frac{\Gamma(\ell+3)}{2\Gamma(\ell-1)}\left(\frac{r}{2M}\right)^{2}\left(1-\frac{2M}{r}\right)~,
\end{align}
which is well-behaved at $r=2M$. While the other branch of the associated Legendre polynomial becomes, 
\begin{align}
Q_{\ell}^{2}(1+2z)\Big|_{\rm near}&=\frac{\sqrt{\pi}}{2^{\ell-1}}\frac{\Gamma(\ell+3)}{2\Gamma(\ell+\frac{3}{2})}\frac{z(1+z)}{(1+2z)^{\ell+3}}
\nonumber
\\
&\times \frac{\Gamma(\ell+\frac{3}{2})\Gamma(-2)}{\Gamma(\frac{\ell}{2}-\frac{1}{2})\Gamma(\frac{\ell}{2})}~,
\end{align}
which is ill-behaved due to the appearance of $\Gamma(-2)$ term. Thus, regularity at the horizon demands $c_{2}=0$ in \ref{staticTeu} and hence the BH TLNs identically vanish. This is consistent with the static limit of the dynamical TLNs associated with the BH spacetimes, as the term which is ill-behaved in the static limit is also ill-behaved on the horizon, and hence does not appear in the BH spacetime. For compact objects other than BHs, on the other hand, one has to impose appropriate boundary conditions, e.g., in the perfectly reflecting case one may impose the condition that the Zerilli and the Regge-Wheeler function should vanish on the surface. Thus, to compare the above result derived solely from the Teukolsky equation in the zero-frequency and zero-rotation limit with those in the literature, we must connect the Teukolsky function with the metric perturbations in the axial and polar sectors. This is what we do next. 

In the non-rotating case, the radial part of the Teukolsky function can be decomposed into axial and polar parts, $R_{\ell}=R_{\ell}^{\rm axial}+R_{\ell}^{\rm polar}$, each of which can be expressed in terms of the Regge-Wheeler and the Zerilli functions respectively, such that in the zero-frequency limit we obtain (see \ref{zero_rot_detweiler} for details), 
\begin{align}
\frac{R_{\ell}^{\textrm{axial}}}{\sqrt{\gamma_{\ell}(2+\gamma_{\ell})}}&=\frac{r^{3}}{8}
\Bigg[V^{\rm axial}\Psi^{\rm RW}_{\ell}+W^{\rm axial}\left(\dfrac{d\Psi_{\ell}^{\rm RW}}{dr_{*}}\right)\Bigg]~,
\\
\frac{R_{\ell}^{\textrm{polar}}}{\sqrt{\gamma_{\ell}(2+\gamma_{\ell})}}&=\frac{r^{3}}{8}
\left[V^{\rm polar}\Psi^{\rm Z}_{\ell}+W^{\rm polar}\left(\dfrac{d\Psi^{\rm Z}_{\ell}}{dr_{*}}\right)\right]~,
\end{align}
where, we have introduced 
\begin{align}
V^{\rm axial}&=V_{\rm RW}=\frac{(r-2M)\left\{(\gamma_{\ell}+2)r-6M\right\}}{r^{4}}~,
\\
W^{\rm axial}&=\frac{2(r-3M)}{r^{2}}~,
\\
V^{\rm polar}&=V_{\rm Z}=\frac{(r-2M)}{r^{4}(\gamma_{\ell}r+6M)^{2}}\Big[\gamma_{\ell}^{2}(\gamma_{\ell}+2)r^{3}
\nonumber
\\
&+6M\gamma_{\ell}^{2}r^{2}+36M^{2}\gamma_{\ell}r+72M^{3}\Big]~,
\\
W^{\rm polar}&=\frac{2\gamma_{\ell}r^{2}-6\gamma_{\ell}Mr-12M^{2}}{r^{2}(\gamma_{\ell}r+6M)}~.
\end{align}
Here, $V_{\rm RW}$ is the Regge-Wheeler and $V_{\rm Z}$ is the Zerilli potential. The above relations connect the radial part of the Teukolsky function with the Regge-Wheeler and the Zerilli functions in the zero-frequency limit. For our purpose, we wish to use the relations connecting the Regge-Wheeler and the Zerilli functions with the axial metric perturbation $h_{0}$ and the polar metric perturbation $H_{0}$, respectively, so that the Teukolsky function can be directly related to the metric perturbations in the zero-frequency case. We start with the polar sector, involving the functions $V^{\rm polar}$ and $W^{\rm polar}$, which can be simplified further as
\begin{widetext}
\begin{align}
V^{\rm polar}\Psi^{\rm Z}_{\ell}&+W^{\rm polar}\left(\dfrac{d\Psi^{\rm Z}_{\ell}}{dr_{*}}\right)
=\frac{(r-2M)}{r^{4}(\gamma_{\ell}r+6M)^{2}}\left[\gamma_{\ell}^{2}(\gamma_{\ell}+2)r^{3}+6M\gamma_{\ell}^{2}r^{2}+36M^{2}\gamma_{\ell}r+72M^{3}\right]\Psi^{\rm Z}_{\ell}
\nonumber
\\
&+\frac{2\gamma_{\ell}r^{2}-6\gamma_{\ell}Mr-12M^{2}}{r^{2}(\gamma_{\ell}r+6M)}\left(\frac{r-2M}{r}\right)\frac{d\Psi^{\rm Z}_{\ell}}{dr}
\nonumber
\\
&=\frac{2(r-2M)}{r^{2}}\Bigg[\frac{\gamma_{\ell}r^{2}-3\gamma_{\ell}Mr-6M^{2}}{r(\gamma_{\ell}r+6M)}
\frac{d\Psi^{\rm Z}_{\ell}}{dr}
+\left(\frac{\gamma_{\ell}^{2}(\frac{\gamma_{\ell}}{2}+1)r^{2}+3M\gamma_{\ell}^{2}r+18M^{2}\gamma_{\ell}+36\frac{M^{3}}{r}}{r(\gamma_{\ell}r+6M)^{2}}\right)\Psi^{\rm Z}_{\ell}\Bigg]
\nonumber
\\
&=\frac{2(r-2M)}{r^{2}}H_{0}~,
\end{align}
\end{widetext}
where, we have used the relation between the polar metric perturbation $H_{0}$ and the Zerilli function $\Psi^{\rm Z}_{\ell}$~\cite{Cardoso:2017cfl}. Therefore, the radial part of the Teukolsky equation associated with the polar perturbation becomes, 
\begin{align}
R_{\ell}^{\rm polar}=\frac{r(r-2M)}{4}\sqrt{\gamma_{\ell}(\gamma_{\ell}+2)}H_{0}~.
\end{align}
In the $z$ coordinate system, the above relation simply becomes $R_{\ell}^{\rm polar}\propto z(1+z)H_{0}$. Thus, from \ref{staticTeu}, we can express the polar metric perturbation $H_{0}$ in terms of the associated Legendre polynomials $P_{\ell}^{2}$, $Q_{\ell}^{2}$, and the arbitrary constants linked with them. Thus, imposing for example Dirichlet boundary conditions on the Zerilli function at the surface of the compact object (note that this is the same boundary condition imposed in~\cite{Cardoso:2017cfl} for a perfectly-reflective mirror), i.e., setting $\Psi_{\ell}^{\rm Z}(r_{*}^{0})=0$, we can determine the ratio of the arbitrary constants, which in turn will yield the TLNs. For the $\ell=2$ mode, we obtain
\begin{align}
k_{2}^{\rm polar}=\frac{8}{5\left(7+3\ln \epsilon \right)}~,
\end{align}
which exactly matches with~\cite{Cardoso:2017cfl}. Thus, the TLNs in the strictly zero-frequency case can indeed be obtained by solving the Teukolsky equation in the non-rotating scenario. Therefore, the formalism developed here can be reconciled with earlier results in the literature~\cite{Cardoso:2017cfl}.

For the axial case, we note that in the zero frequency limit the metric perturbation $h_{0}$ can be expressed as, 
\begin{align}
h_{0}&=\dfrac{d}{dr_{*}}\left(r\Psi_{\ell}^{\rm RW}\right)=r\dfrac{d\Psi_{\ell}^{\rm RW}}{dr_{*}}+\left(\frac{r-2M}{r}\right)\Psi_{\rm RW}~,
\label{h0inRW}
\\
h_{0}'&=\left(\frac{r^{2}}{r-2M}\right)V_{\rm RW}\Psi_{\rm RW}+2\dfrac{d\Psi_{\ell}^{\rm RW}}{dr_{*}}
\nonumber
\\
&\qquad +\Psi_{\ell}^{\rm RW}\left(1+\frac{2M}{r^{2}}\right)~,
\end{align}
where the prime denotes the derivative with respect to the radial coordinate $r$. Given the above expressions for the metric perturbation $h_{0}$ and its derivative $h_{0}'$ in terms of the Regge-Wheeler function and its first derivative, we can consider the following combination $h_{0}'-(2h_{0}/r)$, which does not involve any derivatives of the Regge-Wheeler function, such that,
\begin{align}
\Psi_{\ell}^{\rm RW}=\frac{r^{3}}{\gamma_{\ell}}\dfrac{d}{dr}\left(\frac{h_{0}}{r^{2}}\right)~.
\end{align}
Having expressed the Regge-Wheeler function in terms of $h_{0}$ and its derivative, we can use \ref{h0inRW} to write down the derivative of the Regge-Wheeler function as well in a similar manner, 
\begin{align}
\dfrac{d\Psi_{\ell}^{\rm RW}}{dr_{*}}&=\frac{h_{0}}{r}-\left(\frac{r-2M}{r^{2}}\right)\Psi_{\rm RW}
\nonumber
\\
&=\frac{h_{0}}{r}-\left(\frac{r(r-2M)}{\gamma_{\ell}}\right)\dfrac{d}{dr}\left(\frac{h_{0}}{r^{2}}\right)~.
\end{align}
Such that, we obtain the following combination of the Regge-Wheeler function and its derivative, to read
\begin{widetext}
\begin{align}
V^{\rm axial}\Psi^{\rm RW}_{\ell}&+W^{\rm axial}\left(\dfrac{d\Psi_{\ell}^{\rm RW}}{dr_{*}}\right)
=\frac{(r-2M)\left\{(\gamma_{\ell}+2)r-6M\right\}}{r\gamma_{\ell}}\dfrac{d}{dr}\left(\frac{h_{0}}{r^{2}}\right)
\nonumber
\\
&\qquad \qquad +\frac{2(r-3M)}{r^{2}}\left[\frac{h_{0}}{r}-\left(\frac{r(r-2M)}{\gamma_{\ell}}\right)\dfrac{d}{dr}\left(\frac{h_{0}}{r^{2}}\right)\right]
\nonumber
\\
&=(r-2M)\dfrac{d}{dr}\left(\frac{h_{0}}{r^{2}}\right)+\frac{2(r-3M)}{r^{3}}h_{0}
=\frac{(r-2M)}{r^{2}}\dfrac{dh_{0}}{dr}-\frac{2M}{r^{3}}h_{0}~.
\end{align}
\end{widetext}
Therefore, the axial part of the Teukolsky radial function can be expressed in terms of the metric perturbation $h_{0}$ as,
\begin{align}\label{axial_staticrad}
R_{\ell}^{\textrm{axial}}&=\sqrt{\gamma_{\ell}(2+\gamma_{\ell})}\left[\frac{r(r-2M)}{8}\dfrac{dh_{0}}{dr}-\frac{M}{4}h_{0}\right]~.
\end{align}
The situation for generic choices of $\ell$ is more involved, as the axial metric perturbation $h_{0}$ depends on the MeijerG function, with complicated analytical properties. Thus, we will demonstrate the validity of the above equation for the $\ell=2$ case, which will also be consistent with the results in earlier literature. For the $\ell=2$ mode, the solution of the Teukolsky equation reads 
\begin{align}
R_{\ell}^{\textrm{axial}}&=c_{1}\left(\frac{3r^{2}}{M^{2}}\right)\left(r-2M\right)^{2}
\nonumber
\\
&+c_{2}\left(\frac{r-2M}{M}\right)\left(3r^{2}-6Mr-2M^{2}\right)
\nonumber
\\
&+c_{2}\left(\frac{3r^{2}}{2M^{2}}\right)\left(r-2M\right)^{2}\ln \left(1-\frac{2M}{r}\right)~.
\end{align}
When substituted in \ref{axial_staticrad}, the above solution for the Teukolsky radial function yields the following expression for the axial metric perturbation $h_{0}$, 
\begin{align}
h_{0}&=c_{1}r^{2}(r-2M)+\frac{c_{2}}{24M^{5}r}\Big[4M^{4}+4M^{3}r+6M^{2}r^{2}
\nonumber
\\
&-6Mr^{3}+6Mr^{3}\ln \left(1-\frac{2M}{r}\right)-3r^{4}\ln \left(1-\frac{2M}{r}\right)\Big]~.
\end{align}
One can check that the above expression for the axial metric perturbation $h_{0}$ satisfies the perturbed Einstein's equations in the static limit. Again, given this solution, one can impose Dirichlet boundary condition on the Regge-Wheeler function at the surface of the compact object, by setting $\Psi^{\rm RW}_{2}(r_{*}^{0})=0$. This in turn will relate the two unknown coefficients $c_{1}$ and $c_{2}$ and one arrives at the following expression for the TLN in terms of the compactness parameter $\epsilon$,
\begin{align}
k_{2}^{\rm axial}=\frac{32}{5\left(25+12 \ln \epsilon\right)}~.
\end{align}
Again this coincides with the corresponding expression presented in~\cite{Cardoso:2017cfl} and has the distinctive $\log \epsilon$ structure. Thus, the static limit of the Teukolsky equation and the corresponding solution indeed provide the TLNs with logarithmic dependence, consistent with earlier literature. In summary, the TLNs in the $\omega=0$ case differ from the dynamical TLNs in the $\omega\to 0$ limit in numerical coefficients, though in both cases the logarithmic dependence of the TLNs on $\epsilon$ remains. Thus, the strictly zero frequency case is a set of measure zero, as it corresponds to a single disjoint point in the frequency plane. 

\section{Discussion}\label{sec:conclusions}

TLNs provide a unique opportunity to test the nature of compact objects, as well as the underlying gravitational theory, since only for BHs in GR, the TLNs identically vanish. Earlier results regarding the TLNs of ECOs depicted an astounding behavior, namely the TLNs are non-zero and behave as $\log \epsilon$ in the BH limit, where $\epsilon$ is the difference between the surface of the ECO and the location of the would-be horizon. This suggested that even if $\epsilon$ is small, maybe of the Planckian order, the effect on the TLNs are significant. However, such a result was derived in the context of a static tidal field for non-rotating ECOs, while in practice all the deformations are dynamic and all compact objects are rotating. Thus, it is important to see if such a logarithmic behavior persists even in the dynamical context for rotating ECOs. The inclusion of rotation and dynamics automatically urges one to employ the Teukolsky formalism and it turns out that the Weyl scalars entering the Teukolsky formalism are closely related to the Newtonian potential in the appropriate limits. Thus, the Weyl scalars provide the desired quantity that mimics the role of Newtonian potential in the relativistic context and the Teukolsky equation governs the response function of a compact object under an external tidal field. In this work, we have solved the Teukolsky equation with a generic reflective boundary condition at the surface of the ECO. This in turn leads to a modified tidal response function, and hence on a TLN depending on the reflectivity of the ECO. Since the mode functions are not plane waves in the Teukolsky formalism, we converted the reflectivity in the Teukolsky formalism to the Detweiler formalism using appropriate transformations, and presented the TLNs in terms of the parameters of the ECO and the Detweiler reflectivity, ${\cal R}$.
To our surprise, it turns out that for generic reflectivity the zero-frequency TLNs of a nonspinning compact object identically vanish and non-zero TLNs can be obtained if and only if $\mathcal{R}(\omega\to0)=1$. Intriguingly, the $\log \epsilon$ behavior of the TLNs persists even in the zero-frequency limit of the dynamical TLNs, but it differs in numerical coefficients from the strictly static case and depends on the frequency-dependent part of the reflectivity.
Moreover, as the rotation of the compact object increases the TLNs decrease sharply, implying that fast-rotating objects are difficult to deform by applying external tidal forces. On the other hand, as the reflectivity increases, the TLNs also increase, implying that the more reflective an object is, the higher is its deformation under an external tidal field. Further, we observe certain oscillation patterns in the TLNs, depending on the parameter $\epsilon$. Also, both in the non-rotating and the rotating case, there can be resonances in the TLNs, which happens at higher frequencies for smaller reflectivities. Hence the dynamics of the TLNs are highly non-trivial, as they depend on the parameter $\epsilon$ in an oscillatory manner, while the amplitude of the TLNs depend crucially on the reflectivity and the rotation of the compact object. 

The situation in the zero-frequency limit needs an elaborate discussion. Naively one would expect to arrive at the same result obtained in~\cite{Cardoso:2017cfl} for static perturbations also from the zero-frequency limit of the dynamical TLNs. However, this turns out not to be the case, because the solutions of the Teukolsky equation in the frequency-dependent case become singular in the zero-frequency limit for non-rotating compact objects. Therefore, the $\omega=0$ is an isolated point, which cannot be arrived at by taking $\omega\to 0$, suggesting that the strictly static behavior of the TLNs is a set of measure zero in the phase space.
In other words, the static limit of the dynamical TLNs does \emph{not} coincide with the TLNs derived in the strictly static case.
We would like to emphasize that this behavior has also appeared earlier in the context of the magnetic TLNs of NSs~\cite{Pani:2018inf}, where the $\omega \to 0$ limit of the dynamical TLNs does not coincide with the strictly static TLNs. Indeed, the two types of magnetic TLNs (so-called static and irrotational in that context) have been computed independently in the literature~\cite{Binnington:2009bb,Damour:2009vw}.
As discussed for neutron-star TLNs, the strictly static ones correspond to an idealized situation since the inspiral of a binary system is \emph{always} dynamical~\cite{Landry:2015cva,Pani:2018inf}. Likewise, we consider that the $\omega \to 0$ limit of the dynamical TLNs is the one that should directly affect the gravitational waveform.

We also found that the TLNs generically vanish in the zero-frequency limit, except for the following reflectivity of the compact object: $\mathcal{R}=1+i\omega M\mathcal{R}_{1}+\mathcal{O}(M^{2}\omega^{2})$. Thus, the reflectivity must have a frequency dependence, for the zero-frequency TLNs to be non-zero. 
The condition $\mathcal{R}(\omega\to0)=1$ is generically satisfied by non-dissipative objects, including neutron stars~\cite{MichelaTesi}.
Moreover, the parameter $\mathcal{R}_{1}$ explicitly appears in the reflectivity, and can potentially modify the $\epsilon$-dependence of the TLNs. The $\log \epsilon$ behavior, though generic, can be eliminated by an appropriate choice of $\mathcal{R}_{1}$, the frequency dependent part of the reflectivity. 

Finally, when they are non-zero the TLNs vanish in the BH limit with a typical logarithmic behavior as a function of the distance between the surface of the compact object and the would-be horizon.
This makes the detection of non-zero TLNs for ECOs a tantalizing possibility for future generations of GW detectors~\cite{Cardoso:2017cfl, Maselli:2017cmm}.  

There are several future directions of exploration. For example, whether such behavior for the dynamical TLNs is observed also in the effective field theory description, remains to be seen, see, e.g., two recent papers~\cite{Mandal:2023hqa, Perry:2023wmm}.
The first one claims the existence of logarithmic divergences in the field theory, which must be renormalized to arrive at a finite TLN~\cite{Saketh:2023bul,Mandal:2023hqa}, while the second one arrives at a $\log (r/2M)$ term in the TLN~\cite{Perry:2023wmm}. Both of these logarithmic features are very different from the one discussed here in the context of static TLNs. 
It is also an open point to establish whether the $\omega\to 0$ limit of the dynamical TLNs matches with the strictly static one in the context of effective field theory. A detailed comparison with our results therefore deserves further study.
Also, the decomposition of the TLNs in the axial and polar sectors is not very clear in the context of Weyl scalars, governed by the Teukolsky equation. Furthermore, it could be interesting to reassess the detectability of the TLNs from binary coalescence signals, as the prospects are optimistic owing to the logarithmic scaling~\cite{Cardoso:2017cfl, Maselli:2017cmm}. From the theoretical standpoint, our analysis could be extended to background geometries with fewer symmetries than Kerr~\cite{Raposo:2018rjn}, although that would require going beyond the Teukolsky formalism. Finally, it is possible to map our general framework for TLNs into the membrane paradigm~\cite{Damour:1982, Thorne:1986iy, Price:1986yy}, which can be used to describe ECOs as recently developed in Ref.~\cite{Maggio:2020jml} (see also~\cite{Chen:2020htz, Chakraborty:2022zlq} for recent work on the subject). Work along this direction is underway~\cite{MichelaTesi}.


\begin{acknowledgments}

S.C. acknowledges the warm hospitality at the Albert-Einstein Institute, where a part of this work was performed, which was supported by the Max-Planck-India mobility grant. Research of S.C. is supported by the
MATRICS and the Core research grants from SERB, Government of India (Reg. Nos. MTR/2023/000049
and CRG/2023/000934). 
E.M. is supported by the European Union’s Horizon Europe research and innovation programme under the Marie Skłodowska-Curie grant agreement No. 101107586.
E.M. acknowledges funding from the Deutsche Forschungsgemeinschaft (DFG) - project number: 386119226.
P.P. acknowledges the financial support provided under the European
Union's H2020 ERC, Starting Grant agreement no.~DarkGRA--757480, under
MIUR PRIN (Grant 2020KR4KN2 “String Theory as a bridge between Gauge Theories and Quantum Gravity”) and FARE (GW-NEXT, CUP: B84I20000100001, 2020KR4KN2) programs, and support from the Amaldi Research Center funded by the MIUR program ``Dipartimento di Eccellenza" (CUP:~B81I18001170001).
\end{acknowledgments}
\appendix
\labelformat{section}{Appendix #1} 
\labelformat{subsection}{Appendix #1}
\begin{widetext}
\section{Connecting the near-horizon behavior of the Teukolsky function to the Detweiler variable}
\label{TeuDet}

The first step of connecting the near-horizon behavior of the radial Teukolsky function in the $(v,r,\theta,\tilde{\phi})$ coordinate with the Detweiler function, is to transform the radial Teukolsky function in the Boyer-Lindquist coordinate system. For that the radial function in the Boyer-Lindquist coordinate will have an additional factor of $e^{-i\omega r_{*}}$ (as, $v=t+r_{*}$) and also a factor of $e^{im\bar{r}_{*}}$ (as, $\tilde{\phi}=\phi+\bar{r}_{*}$) in the near horizon limit. Therefore the radial Teukolsky function in the $(t,r,\theta,\phi)$ coordinate system becomes,
\begin{align}
\,_{-2}R_{\ell m}^{(t)}=\mathcal{A}\frac{\Delta^{2}}{(r_{+}-r_{-})^{4}}e^{-i\omega r_{*}}e^{im\bar{r}_{*}}+\mathcal{B}e^{-i\omega r_{*}}e^{im\bar{r}_{*}}e^{2i\bar{\omega}r_{*}}
\end{align}
where the tortoise coordinate $r_{*}$ and the azimuthal tortoise coordinate $\bar{r}_{*}$ are given by, 
\begin{align}
\frac{dr_{*}}{dr}=\frac{r^{2}+a^{2}}{\Delta}~;\qquad \frac{d\bar{r}_{*}}{dr}=\frac{a}{\Delta}~.
\end{align}
In the near horizon limit, both of these tortoise coordinates get related to each other, such that, 
\begin{align}
d\bar{r}_{*}=\frac{a}{\Delta}dr=\frac{a}{r_{+}^{2}+a^{2}}\frac{r_{+}^{2}+a^{2}}{\Delta}dr=\frac{a}{r_{+}^{2}+a^{2}}dr_{*}~.
\end{align}
Hence the radial Teukolsky function in the Boyer-Lindquist coordinate system can expressed solely in terms of the tortoise coordinate $r_{*}$ and hence the radial function becomes, 
\begin{align}
\,_{-2}R_{\ell m}^{(t)}=\mathcal{A}\frac{\Delta^{2}}{(r_{+}-r_{-})^{4}}e^{-i\bar{\omega} r_{*}}+\mathcal{B}e^{i\bar{\omega} r_{*}}~,
\end{align}
where, 
\begin{align}
\bar{\omega}=\omega-\frac{am}{r_{+}^{2}+a^{2}}=\omega-m\Omega_{\rm H}~.
\end{align}
Here, $\Omega_{\rm H}$ is the angular velocity of the horizon. The above expression for the radial Teukolsky function in the Boyer-Lindquist coordinate has been used in the main text. 

The next task is to transform the near-horizon radial Teukolsky function in the Boyer-Lindquist coordinate to Detweiler function, through~\ref{Teu_Det_Trans}. Since the transformation is singular on the horizon, the near horizon limit must be taken with care. For this purpose, we expand the quantities $\mathcal{A}$ and $\mathcal{B}$ in the near horizon regime as, $\mathcal{A}=A_{0}$ and $\mathcal{B}=B_{0}+B_{1}\Delta+B_{2}\Delta^{2}$, where $A_{0}$, $B_{0}$, $B_{1}$ and $B_{2}$ are constants, independent of the radial coordinate. Thus, the radial Teukolsky function in the Boyer-Lindquist coordinate becomes
\begin{align}\label{expand_teu}
\,_{-2}R_{\ell m}^{(t)}=A_{0}\frac{\Delta^{2}}{(r_{+}-r_{-})^{4}}e^{-i\bar{\omega} r_{*}}+\left(B_{0}+B_{1}\Delta+B_{2}\Delta^{2}\right)e^{i\bar{\omega} r_{*}}~.
\end{align}
Of course, the quantities $A_{0}$, $B_{0}$, $B_{1}$ and $B_{2}$ cannot be arbitrary, rather they will be fixed by the Teukolsky equation. One substitutes the above expression for the radial perturbation in the radial Teukolsky equation and demands that the radial Teukolsky equation will be satisfied at all orders in $\Delta$, thereby uniquely fixing the above quantities. In what follows, we will choose these quantities to have precisely those values, which are consistent with the radial Teukolsky equation.   

Moving forward, the determination of the Detweiler function requires determining the derivative of the radial Teukolsky function with respect to $r$ in the Boyer-Lindquist coordinate, which yields, 
\begin{align}
\dfrac{d}{dr}\left(_{-2}R^{(t)}_{\ell m}\right)&=A_{0}\frac{\Delta^{2}}{(r_{+}-r_{-})^{4}}e^{-i\bar{\omega} r_{*}}\left(-\frac{i\bar{\omega}(r_{+}^{2}+a^{2})}{\Delta} \right)+A_{0}\frac{2\Delta}{(r_{+}-r_{-})^{3}}e^{-i\bar{\omega} r_{*}}
\nonumber
\\
&+\left(B_{0}+B_{1}\Delta+B_{2}\Delta^{2}\right)e^{i\bar{\omega} r_{*}}\left(\frac{i\bar{\omega}(r_{+}^{2}+a^{2})}{\Delta} \right)+\left\{B_{1}(r_{+}-r_{-})+2B_{2}\Delta(r_{+}-r_{-})\right\}e^{i\bar{\omega} r_{*}}~.
\end{align}
Substituting the above expression for the radial derivative of the radial Teukolsky function, along with the radial Teukolsky function itself in~\ref{Teu_Det_Trans}, we obtain the following expression for the Detweiler function, 
\begin{align}
\,_{-2}X_{\ell m}&=\sqrt{r_{+}^{2}+a^{2}}\left[\frac{\alpha}{\Delta}R_{-2}^{(t)}+\frac{\beta}{\Delta^{2}}\dfrac{dR_{-2}^{(t)}}{dr}\right]
\nonumber
\\
&=\sqrt{r_{+}^{2}+a^{2}}\Bigg[\frac{\alpha}{\Delta}\left\{A_{0}\frac{\Delta^{2}}{(r_{+}-r_{-})^{4}}e^{-i\bar{\omega} r_{*}}+\left(B_{0}+B_{1}\Delta+B_{2}\Delta^{2}\right)e^{i\bar{\omega} r_{*}} \right\}
\nonumber
\\
&\quad +\frac{\beta}{\Delta^{2}}\Bigg\{ A_{0}\frac{\Delta^{2}}{(r_{+}-r_{-})^{4}}e^{-i\bar{\omega} r_{*}}\left(-\frac{i\bar{\omega}(r_{+}^{2}+a^{2})}{\Delta} \right)+A_{0}\frac{2\Delta}{(r_{+}-r_{-})^{3}}e^{-i\bar{\omega} r_{*}}
\nonumber
\\
&\quad +\left(B_{0}+B_{1}\Delta+B_{2}\Delta^{2}\right)e^{i\bar{\omega} r_{*}}\left(\frac{i\bar{\omega}(r_{+}^{2}+a^{2})}{\Delta} \right)+\left\{B_{1}(r_{+}-r_{-})+2B_{2}\Delta(r_{+}-r_{-})\right\}e^{i\bar{\omega} r_{*}}\Bigg\}\Bigg]
\nonumber
\\
&=\sqrt{r_{+}^{2}+a^{2}}\left[\frac{\alpha \Delta}{(r_{+}-r_{-})^{4}}+\frac{2\beta}{\Delta(r_{+}-r_{-})^{3}}-\frac{\beta}{\Delta}\left(\frac{i\bar{\omega}(r_{+}^{2}+a^{2})}{(r_{+}-r_{-})^{4}} \right) \right]A_{0}e^{-i\bar{\omega} r_{*}}
\nonumber
\\
&\quad +\sqrt{r_{+}^{2}+a^{2}}\Bigg[B_{0}\frac{\alpha}{\Delta}+B_{1}\alpha+B_{2}\alpha\Delta+B_{1}(r_{+}-r_{-})\frac{\beta}{\Delta^{2}}+2B_{2}(r_{+}-r_{-})\frac{\beta}{\Delta}
\nonumber
\\
&\quad +\left(B_{0}+B_{1}\Delta+B_{2}\Delta^{2}\right)\left(\frac{i\beta\bar{\omega}(r_{+}^{2}+a^{2})}{\Delta^{3}} \right)
\Bigg]e^{i\bar{\omega} r_{*}}
\nonumber
\\
&\equiv \mathcal{A}_{X}e^{-i\bar{\omega} r_{*}}+\mathcal{B}_{X}e^{i\bar{\omega} r_{*}}~,
\end{align}
where the two radial coordinate dependent quantities, namely $\mathcal{A}_{X}$ and $\mathcal{B}_{X}$, sitting in front of the ingoing and outgoing solution near the horizon can be expressed as, 
\begin{align}
\mathcal{A}_{X}&=\frac{\sqrt{r_{+}^{2}+a^{2}}}{(r_{+}-r_{-})^{4}}\left[\alpha \Delta+\left\{2(r_{+}-r_{-})-i\bar{\omega}(r_{+}^{2}+a^{2})\right\}\frac{\beta}{\Delta}\right]A_{0}
\\
\mathcal{B}_{X}&=\sqrt{r_{+}^{2}+a^{2}}\Bigg[B_{0}\left\{\frac{\alpha}{\Delta}+i\bar{\omega}(r_{+}^{2}+a^{2})\frac{\beta}{\Delta^{3}} \right\}+B_{1}\left\{\alpha+(r_{+}-r_{-})\frac{\beta}{\Delta^{2}}+i\bar{\omega}(r_{+}^{2}+a^{2})\frac{\beta}{\Delta^{2}} \right\}
\nonumber
\\
&\qquad+B_{2}\left\{\alpha\Delta+2(r_{+}-r_{-})\frac{\beta}{\Delta}+i\bar{\omega}(r_{+}^{2}+a^{2})\frac{\beta}{\Delta}\right\}\Bigg]~.
\end{align}
Thus, the reflectivity of the compact object, whose surface is very close to the BH horizon, can be determined by taking the ratio $(\mathcal{B}_{X}/\mathcal{A}_{X})$, which reads,
\begin{align}
\mathcal{R}&=\lim_{r\rightarrow r_{0}}(r_{+}-r_{-})^{4}\left[\alpha \Delta+\left\{2(r_{+}-r_{-})-i\bar{\omega}(r_{+}^{2}+a^{2})\right\}\frac{\beta}{\Delta}\right]^{-1}\Bigg\{\Bigg[\frac{B_{0}}{A_{0}}\left\{\frac{\alpha}{\Delta}+i\bar{\omega}(r_{+}^{2}+a^{2})\frac{\beta}{\Delta^{3}} \right\}
\nonumber
\\
&\qquad+\frac{B_{1}}{A_{0}}\left\{\alpha+(r_{+}-r_{-})\frac{\beta}{\Delta^{2}}+i\bar{\omega}(r_{+}^{2}+a^{2})\frac{\beta}{\Delta^{2}} \right\}
+\frac{B_{2}}{A_{0}}\left\{\alpha\Delta+2(r_{+}-r_{-})\frac{\beta}{\Delta}+i\bar{\omega}(r_{+}^{2}+a^{2})\frac{\beta}{\Delta}\right\}\Bigg]\Bigg\}~.
\end{align}
Note that the coefficients $A_{0}$, $A_{1}$, $B_{0}$, $B_{1}$ and $B_{2}$, appearing in the above expression for the Detweiler reflectivity, are to be fixed by demanding that the radial function with the above coefficients, as in \ref{expand_teu}, follows the Teukolsky equation. Moreover, the reflectivity depends on the quantities $\alpha$ and $\beta$ appearing in the transformation from the Teukolsky to the Detweiler functions and have very specific behavior near the horizon, which must also be taken into account. All in all, the above limit needs to be taken carefully, since there can be various divergent contributions canceling each other, yielding a finite value for the reflectivity $\mathcal{R}$ as a function of the object parameters, frequency, the ratio $\mathcal{B}/\mathcal{A}$ and of course $\epsilon$. This is the result we have used in the main text. 

\section{Analytical expression for the TLN in the small-frequency limit}\label{app:smallfrequency}

In this appendix, we provide the derivation of the analytical expression for the TLN for rotating spacetimes in the small-frequency regime, i.e., $M\omega \ll 1$. The starting point is the radial Teukolsky equation in the $(v,r,\theta,\phi)$ coordinate, as presented in~\ref{Teukeq}. As emphasized in the main text, besides the small frequency approximation, we also require the near-zone limit ($\omega r\ll 1$) in order to arrive at an analytical expression for the frequency-dependent TLN. The near-zone expansion of~\ref{Teukeq} at the leading order in the frequency reads~\cite{Chia:2020yla}
\begin{align}
\dfrac{d^{2}\,_{-2}R^{(v)}_{\ell m}}{dz^{2}}+\left[\frac{2iP_{+}-1}{z}-\frac{2iP_{+}+1}{z+1} \right]\dfrac{d\,_{-2}R^{(v)}_{\ell m}}{dz}+\left[\frac{4iP_{+}}{(1+z)^{2}}-\frac{4iP_{+}}{z^{2}}-\frac{\ell(\ell+1)-2}{z(1+z)}\right]\,_{-2}R^{(v)}_{\ell m}=0~,
\end{align}
where we have defined the quantity $z$ as,
\begin{align}\label{defz}
z\equiv \frac{r-r_{+}}{r_{+}-r_{-}}~;\qquad P_{+}=\frac{am-2M\omega r_{+}}{r_{+}-r_{-}}=-\frac{2Mr_{+}\bar{\omega}}{r_{+}-r_{-}}~,
\end{align}
and the expression for $P_{+}$ follows from~\ref{Pplusminus}. Further, we have introduced the following definition for rescaled frequency $\bar{\omega}$ as, $\bar{\omega}\equiv\omega-m\Omega_{\rm H}$, where $\Omega_{\rm H}$ is the angular velocity of the horizon. The above equation can be explicitly solved in terms of the Hypergeometric functions, yielding, 
\begin{align}\label{gen_sol}
\,_{-2}R^{(v)}_{\ell m}&=\mathcal{A}z^{2}(1+z)^{2}~\,_{2}F_{1}\left(2-\ell,3+\ell,3+2iP_{+};-z\right)
\nonumber
\\
&\qquad \qquad +\mathcal{B}z^{-2iP_{+}}(1+z)^{2}~\,_{2}F_{1}\left(-\ell-2iP_{+},1+\ell-2iP_{+},-1-2iP_{+};-z\right)~,
\end{align}
where, $\mathcal{A}$ and $\mathcal{B}$ are the arbitrary constants of integration. Therefore, close to the horizon we can use the property that $\lim_{x\rightarrow 0}\,_{2}F_{1}(a,b;c;x)=1$, in order to obtain the following expression for the radial Teukolsky function:
\begin{align}
\,_{-2}R^{(v)}_{\ell m}\Big|_{\rm near-horizon}&=\mathcal{A}z^{2}+\mathcal{B}z^{-2iP_{+}}
\nonumber
\\
&=\frac{\mathcal{A}}{(r_{+}-r_{-})^{4}}\Delta^{2}+\mathcal{B}\exp\left[2i\left(\frac{r_{+}^{2}+a^{2}}{r_{+}-r_{-}}\right)\bar{\omega}\ln z \right]~.
\label{radialvnear}
\end{align}
In order to arrive at the first term in the second line, we have used the relation between $z$ and the radial coordinate $r$, while the second term has been obtained by relating the tortoise coordinate $r_{*}$ to the near-horizon coordinate $z$. The determination of the arbitrary constants $\mathcal{A}$ and $\mathcal{B}$ are to be performed by imposing appropriate boundary conditions on the radial Teukolsky wave function close to the horizon. 

The asymptotic limit of the hypergeometric functions, on the other hand, can be used to determine the radial part of the rescaled and perturbed Weyl scalar $\rho^{4}\Psi_{4}$ in the intermediate regime. This region is far from the surface of the compact object ($r\gg r_{+}$) but is much closer to the compact object than the distance of the compact object from the source of the tidal field ($r\ll b$, where $b$ is the distance between the compact object and the source of the tidal field) and is contained within the near-zone region. In this region, the radial part of the perturbed and rescaled Weyl scalar, obtained in~\ref{gen_sol}, becomes,
\begin{align}
\,_{-2}R^{(v)}_{\ell m}\Big|_{\rm far}&=\mathcal{A}z^{4}\left\{\frac{\Gamma(3+2iP_{+})\Gamma(1+2\ell)}{\Gamma(3+\ell)\Gamma(1+\ell+2iP_{+})}z^{\ell-2}+\frac{\Gamma(3+2iP_{+})\Gamma(-1-2\ell)}{\Gamma(2-\ell)\Gamma(-\ell+2iP_{+})}z^{-\ell-3}\right\}
\nonumber
\\
&\qquad \qquad + \mathcal{B}z^{2-2iP_{+}}\left\{\frac{\Gamma(-1-2iP_{+})\Gamma(1+2\ell)}{\Gamma(1+\ell-2iP_{+})\Gamma(\ell-1)}z^{\ell+2iP_{+}}+\frac{\Gamma(-1-2iP_{+})\Gamma(-1-2\ell)}{\Gamma(-\ell-2iP_{+})\Gamma(-2-\ell)}z^{2iP_{+}-1-\ell}\right\}
\nonumber
\\
&=z^{4}\Bigg[\left\{\mathcal{A}\frac{\Gamma(3+2iP_{+})\Gamma(1+2\ell)}{\Gamma(3+\ell)\Gamma(1+\ell+2iP_{+})}+\mathcal{B}\frac{\Gamma(-1-2iP_{+})\Gamma(1+2\ell)}{\Gamma(1+\ell-2iP_{+})\Gamma(\ell-1)}\right\}z^{\ell-2}
\nonumber
\\
&\qquad \qquad +\left\{\mathcal{A}\frac{\Gamma(3+2iP_{+})\Gamma(-1-2\ell)}{\Gamma(2-\ell)\Gamma(-\ell+2iP_{+})}+\mathcal{B}\frac{\Gamma(-1-2iP_{+})\Gamma(-1-2\ell)}{\Gamma(-\ell-2iP_{+})\Gamma(-2-\ell)} \right\} z^{-\ell-3}\Bigg]
\nonumber
\\
&=z^{\ell+2}\left\{\mathcal{A}\frac{\Gamma(3+2iP_{+})\Gamma(1+2\ell)}{\Gamma(3+\ell)\Gamma(1+\ell+2iP_{+})}+\mathcal{B}\frac{\Gamma(-1-2iP_{+})\Gamma(1+2\ell)}{\Gamma(1+\ell-2iP_{+})\Gamma(\ell-1)}\right\}\Bigg[1+\mathcal{F}_{\ell}z^{-2\ell-1} \Bigg]~.
\end{align}
A comparison of $\,_{-2}R^{(v)}_{\ell m}$, derived above, depicting the radial part of the rescaled and perturbed Weyl scalar $\rho^{4}\Psi_{4}$ in the intermediate zone, with~\ref{limpsi4} reveals that one can identify the tidal response function $\mathcal{F}_{\ell}$ (note that the response function being independent of the azimuthal number $m$, we have omitted it from the subscript) to have the following expression (see also~\cite{Nair:2022xfm})
\begin{align}\label{flm}
\mathcal{F}_{\ell}&=\frac{\mathcal{A}\frac{\Gamma(3+2iP_{+})\Gamma(-1-2\ell)}{\Gamma(2-\ell)\Gamma(-\ell+2iP_{+})}+\mathcal{B}\frac{\Gamma(-1-2iP_{+})\Gamma(-1-2\ell)}{\Gamma(-\ell-2iP_{+})\Gamma(-2-\ell)}}{\mathcal{A}\frac{\Gamma(3+2iP_{+})\Gamma(1+2\ell)}{\Gamma(3+\ell)\Gamma(1+\ell+2iP_{+})}+\mathcal{B}\frac{\Gamma(-1-2iP_{+})\Gamma(1+2\ell)}{\Gamma(1+\ell-2iP_{+})\Gamma(\ell-1)}}
\nonumber
\\
&=\frac{\Gamma(3+\ell)\Gamma(1+\ell+2iP_{+})\Gamma(-1-2\ell)}{\Gamma(2-\ell)\Gamma(-\ell+2iP_{+})\Gamma(1+2\ell)}\left[\frac{1+\frac{\mathcal{B}}{\mathcal{A}}\Gamma_{2}}{1+\frac{\mathcal{B}}{\mathcal{A}}\Gamma_{1}} \right]~,
\end{align}
where, we have introduced two quantities $\Gamma_{1}$ and $\Gamma_{2}$, dependent on the frequency and the angular number $\ell$ as,
\begin{align}
\Gamma_{1}&=\frac{\Gamma(-1-2iP_{+})\Gamma(3+\ell)\Gamma(1+\ell+2iP_{+})}{\Gamma(1+\ell-2iP_{+})\Gamma(3+2iP_{+})\Gamma(\ell-1)}~,
\\
\Gamma_{2}&=\frac{\Gamma(-1-2iP_{+})\Gamma(2-\ell)\Gamma(-\ell+2iP_{+})}{\Gamma(-\ell-2iP_{+})\Gamma(3+2iP_{+})\Gamma(-2-\ell)}~.
\end{align}
The real part of the tidal response $\mathcal{F}_{\ell}$ provides the TLN. Note that the above expression for the tidal response function can be further simplified, along with the quantities $\Gamma_{1}$ and $\Gamma_{2}$, using various identities involving Gamma functions with imaginary arguments. In particular, we can express the term outside the square bracket in~\ref{flm} as
\begin{align}
\frac{\Gamma(3+\ell)\Gamma(1+\ell+2iP_{+})\Gamma(-1-2\ell)}{\Gamma(2-\ell)\Gamma(-\ell+2iP_{+})\Gamma(1+2\ell)}
&=\frac{\Gamma(3+\ell)\Gamma(1+\ell+2iP_{+})\Gamma(1+\ell-2iP_{+})\Gamma(-1-2\ell)}{\Gamma(2-\ell)\Gamma(-\ell+2iP_{+})\Gamma(1+\ell-2iP_{+})\Gamma(1+2\ell)}
\nonumber
\\
&=\frac{\Gamma(3+\ell)\Gamma(-1-2\ell)}{\Gamma(2-\ell)\Gamma(1+2\ell)}\left[\frac{2\pi P_{+}}{\sinh(2\pi P_{+})}\prod_{j=1}^{\ell}\left(j^{2}+4P_{+}^{2}\right)\right]\left[\frac{\sin\left(-\pi \ell+2i\pi P_{+} \right)}{\pi}\right]
\nonumber
\\
&=\left(-1\right)^{\ell}\frac{\Gamma(3+\ell)\Gamma(1+\ell)\Gamma(-1-2\ell)\Gamma(2+2\ell)}{\Gamma(1+\ell)\Gamma(-\ell)(\ell-1)\ell\Gamma(1+2\ell)\Gamma(2+2\ell)}\left[2iP_{+}\prod_{j=1}^{\ell}\left(j^{2}+4P_{+}^{2}\right)\right]
\nonumber
\\
&=\left(-1\right)^{\ell}\frac{\pi}{\sin(2\pi+2\pi\ell)}\left(\frac{(\ell+2)!(\ell-2)!}{(2\ell)!(1+2\ell)!}\right)\frac{\sin(\pi+\pi \ell)}{\pi}\left[2iP_{+}\prod_{j=1}^{\ell}\left(j^{2}+4P_{+}^{2}\right)\right]
\nonumber
\\
&=-iP_{+}\left(\frac{(\ell+2)!(\ell-2)!}{(2\ell)!(1+2\ell)!}\right)\prod_{j=1}^{\ell}\left(j^{2}+4P_{+}^{2}\right)
\label{overall_factor}
\end{align}
where, along with the results --- (a) $\sin(ix)=i\sinh(x)$, (b) $\sin(2x)=2\sin(x)\cos(x)$, and (c) $\Gamma(1+n)=n!$, we have also used the following identities involving $\Gamma$ functions: 
\begin{align}
\Gamma(z)\Gamma(1-z)&=\frac{\pi}{\sin (\pi z)}~,
\\
|\Gamma(1+n+ib)|^{2}&=\frac{\pi b}{\sinh(\pi b)}\prod_{j=1}^{n}\left(j^{2}+b^{2}\right)~;
\qquad j\in \mathbb{N} \,.
\end{align}
The functions $\Gamma_{1}$ and $\Gamma_{2}$ can also be expressed as,
\begin{align}
\Gamma_{1}&=\frac{(\ell+2)!}{(\ell-2)!}\frac{\Gamma(-1-2iP_{+})\Gamma(1+\ell+2iP_{+})}{\Gamma(1+\ell-2iP_{+})\Gamma(3+2iP_{+})}
\nonumber
\\
&=\frac{(\ell+2)!}{(\ell-2)!}\frac{\left(\ell+2iP_{+}\right)\left(\ell-1+2iP_{+}\right)\times \cdots \times\left(3+2iP_{+}\right)}{\left(\ell-2iP_{+}\right)\left(\ell-1-2iP_{+}\right)\times \cdots \times \left(-2iP_{+}\right) \left(-1-2iP_{+}\right)} \,,
\label{Gamma_1}
\\
\Gamma_{2}&=\frac{\Gamma(-1-2iP_{+})\Gamma(2-\ell)\Gamma(\ell-1)\Gamma(-\ell+2iP_{+})}{\Gamma(-\ell-2iP_{+})\Gamma(3+2iP_{+})\Gamma(-2-\ell)\Gamma(3+\ell)}\frac{\Gamma(3+\ell)}{\Gamma(\ell-1)}
\nonumber
\\
&=\frac{(\ell+2)!}{(\ell-2)!}\frac{\pi}{\sin \left[\pi(\ell-1)\right]}\frac{\sin \left[\pi(\ell+3) \right]}{\pi}\frac{\Gamma(-1-2iP_{+})\Gamma(-\ell+2iP_{+})}{\Gamma(-\ell-2iP_{+})\Gamma(3+2iP_{+})}
\nonumber
\\
&=\frac{(\ell+2)!}{(\ell-2)!}
\frac{\left(-2-2iP_{+}\right)\left(-3-2iP_{+}\right)\times \cdots \times\left(-\ell-2iP_{+}\right)}{\left(2+2iP_{+}\right)\left(1+2iP_{+}\right)\left(2iP_{+}\right)\times \cdots \times\left(-\ell+2iP_{+}\right)}
\nonumber
\\
&=\frac{(\ell+2)!}{(\ell-2)!}
\frac{(-1)^{\ell-1}\left(2+2iP_{+}\right)\left(3+2iP_{+}\right)\times \cdots \times\left(\ell+2iP_{+}\right)}{(-1)^{\ell+2}\left(2+2iP_{+}\right)\left(-1-2iP_{+}\right)\left(-2iP_{+}\right)\times \cdots \times\left(\ell-2iP_{+}\right)}
=-\Gamma_{1} \,,
\label{Gamma_2}
\end{align}
where, we have used the result, $\sin[\pi(3+\ell)]=\sin[\pi(\ell-1)]\cos (4\pi)+\sin[4\pi]\cos[\pi(\ell-1)]=\sin[\pi(\ell-1)]$. Therefore, the tidal response function depends on $\Gamma_{1}$ alone. This result, along with the expression for $\Gamma_{1}$ has been presented in the main text. 

Having discussed the most general situation with arbitrary choices for $\ell$ and $a$, let us briefly touch upon the interesting limit, with $\ell=2$ and $a=0$. In this case, it follows that $P_{+}=-2M\omega$ and the expression for $\Gamma_{1}$ simplifies to, 
\begin{align}
\Gamma_{1}&=24\frac{\Gamma(-1+4iM\omega)}{\Gamma(3+4iM\omega)}=\frac{24}{(2+4iM\omega)(1+4iM\omega)4iM\omega(-1+4iM\omega)}\simeq \frac{3i}{M\omega}~.
\end{align}
Moreover, we also obtain, 
\begin{align}
\Gamma_{2}=24\frac{\Gamma(-1+4iM\omega)\Gamma(-2-4iM\omega)}{\Gamma(-2+4iM\omega)\Gamma(3-4iM\omega)}
\simeq -\frac{3i}{M\omega}~.
\end{align}
Thus, in the non-rotating case, with $\ell=2$, we obtain the following expression for the response function 
\begin{align}
F_{2}&=\frac{8iM\omega}{5!}\left(1+16M^{2}\omega^{2}\right)\left(1+4M^{2}\omega^{2}\right)
\frac{1+24\left(\frac{\mathcal{B}}{\mathcal{A}}\right)\frac{\Gamma(-1+4iM\omega)\Gamma(-2-4iM\omega)}{\Gamma(-2+4iM\omega)\Gamma(3-4iM\omega)}}{1+24\left(\frac{\mathcal{B}}{\mathcal{A}}\right)\frac{\Gamma(-1+4iM\omega)}{\Gamma(3+4iM\omega)}}
\nonumber
\\
&=\frac{8iM\omega}{5!}\left(1+16M^{2}\omega^{2}\right)\left(1+4M^{2}\omega^{2}\right)\frac{\Gamma(3+4iM\omega)}{\Gamma(-2+4iM\omega)\Gamma(3-4iM\omega)}
\nonumber
\\
&\qquad \times\Bigg[\frac{\Gamma(-2+4iM\omega)\Gamma(3-4iM\omega)+24\left(\frac{\mathcal{B}}{\mathcal{A}}\right)\Gamma(-1+4iM\omega)\Gamma(-2-4iM\omega)}{\Gamma(3+4iM\omega)+24\left(\frac{\mathcal{B}}{\mathcal{A}}\right)\Gamma(-1+4iM\omega)}\Bigg]~.
\end{align}
Further simplification can be achieved by using the following identity, 
\begin{align}
\Gamma(3-4iM\omega)&=(2-4iM\omega)(1-4iM\omega)(-4iM\omega)(-1-4iM\omega)(-2-4iM\omega)\Gamma(-2-4iM\omega)
\nonumber
\\
&=-16iM\omega\left(1+16M^{2}\omega^{2}\right)\left(1+4M^{2}\omega^{2}\right)\Gamma(-2-4iM\omega)~,
\end{align}
so that the response function becomes, 
\begin{align}\label{TeuF2}
F_{2}&=-\frac{1}{240}\frac{\Gamma(3+4iM\omega)}{\Gamma(-2+4iM\omega)\Gamma(-2-4iM\omega)}
\nonumber
\\
&\qquad\qquad\qquad\times\Bigg[\frac{\Gamma(-2+4iM\omega)\Gamma(3-4iM\omega)+24\left(\frac{\mathcal{B}}{\mathcal{A}}\right)\Gamma(-1+4iM\omega)\Gamma(-2-4iM\omega)}{\Gamma(3+4iM\omega)+24\left(\frac{\mathcal{B}}{\mathcal{A}}\right)\Gamma(-1+4iM\omega)}\Bigg]~.
\end{align}
Note that the above expression is in terms of the Teukolsky reflectivity $(\mathcal{B}/\mathcal{A})$, which needs to be converted to the Detweiler reflectivity $\mathcal{R}$. The steps involved in performing this modification is detailed in~\ref{TeuDet}, while here we simply quote the final expression for the response function of the $\ell=2$ mode, in the non-rotating limit, in terms of the Detweiler reflectivity,   
\begin{align}\label{DetF2}
F_{2}&=-\frac{1}{240}\frac{\Gamma(3+4iM\omega)}{\Gamma(-2+4iM\omega)\Gamma(-2-4iM\omega)}
\nonumber
\\
&\qquad\qquad\qquad\times \Bigg[\frac{e^{4iM\omega(\epsilon +\ln \epsilon)}(2-iM\omega)\Gamma(-2+4iM\omega)\Gamma(3-4iM\omega)+\widetilde{\mathcal{R}}\Gamma(-1+4iM\omega)\Gamma(-2-4iM\omega)}{e^{4iM\omega(\epsilon +\ln \epsilon)}\Gamma(2-iM\omega)\Gamma(3+4iM\omega)+\widetilde{\mathcal{R}}\Gamma(-1+4iM\omega)}\Bigg]~,
\end{align}
where the frequency dependent quantity $\widetilde{\mathcal{R}}$ is related to the Detweiler reflectivity $\mathcal{R}$ through the following relation, 
\begin{align}
\widetilde{\mathcal{R}}=16e^{8\pi M\omega}M\omega(i+2M\omega+16iM^{2}\omega^{2}+32M^{3}\omega^{3}) \mathcal{R}~.
\end{align}
Therefore, a comparison between the response function in~\ref{TeuF2}, obtained by using the Teukolsky function, and the one in~\ref{DetF2}, determined by the Detweiler function, reveals the following connection between the two reflectivities,   
\begin{align}
\frac{\mathcal{B}}{\mathcal{A}}&=\frac{2}{3}\frac{e^{8\pi M\omega}M\omega \mathcal{R}(i+2M\omega+16iM^{2}\omega^{2}+32M^{3}\omega^{3})}{e^{4iM\epsilon \omega}\epsilon^{4iM\omega}(2-iM\omega)}
\nonumber
\\
&=e^{8\pi M\omega-4iM\omega\left(\epsilon+\ln \epsilon\right)}\frac{M\omega}{3}\frac{(i+2M\omega+16iM^{2}\omega^{2}+32M^{3}\omega^{3})}{1-\frac{iM\omega}{2}}\mathcal{R}~.
\end{align}
Note that, the tortoise coordinate $r^{*}_{0}$ at the location of the reflective surface can be expressed in terms of the parameter $\epsilon$, which depicts the shift of the reflective surface from the would-be horizon as
\begin{align}
r^{*}_{0}=2M\left(1+\epsilon\right)+2M\ln \epsilon=2M\left(1+\epsilon+\ln \epsilon\right)~,
\end{align}
and hence the relation between the Teukolsky and the Detweiler reflectivity takes the following form,
\begin{align}
\frac{\mathcal{B}}{\mathcal{A}}=e^{8\pi M\omega-2i\omega\left(r^{*}_{0}-2M\right)}\frac{(i+2M\omega+16iM^{2}\omega^{2}+32M^{3}\omega^{3})}{1-\frac{iM\omega}{2}}\left(\frac{M\omega \mathcal{R}}{3}\right)~.
\end{align}
In the limit of $\omega\rightarrow 0$, the Teukolsky and the Detweiler reflectivity gets related in a simple form, which reads, $(\mathcal{B}/\mathcal{A})=(iM\omega \mathcal{R}/3)$. We will use this particular form in the main text while discussing the static limit of the TLN for a non-rotating BH. 

\section{Teukolsky equation with quadratic-in-frequency corrections}\label{quad_freq}

We start by simplifying \ref{Teukeq} by keeping terms up to $\mathcal{O}(M^{2}\omega^{2})$. For this purpose, we note the following identities among the parameters of \ref{Teukeq}, 
\begin{align}
P_{+}-P_{-}&=-2M\omega~,\qquad P_{-}=P_{+}+2M\omega~,
\\
A_{+}-A_{-}&=-2(r_{+}-r_{-})(P_{+}-P_{-})\omega-(r_{+}+2M)r_{+}\omega^{2}+(r_{-}+2M)r_{-}\omega^{2}
\nonumber
\\
&=-\omega^{2}(r_{+}-r_{-})(r_{+}+r_{-}-2M)~,
\\
B_{+}-B_{-}&=2\omega(r_{+}-r_{-})~.
\end{align}
Besides, note that $E_{\ell m}=\ell(\ell+1)+E_{1}a\omega+E_{2}a^{2}\omega^{2}$, where $E_{1}$ and $E_{2}$ depends on the angular and azimuthal numbers $\ell$ and $m$, along with the mass of the BH. Changing the radial coordinate $r$ to a new coordinate $z$, defined in \ref{defz}, and using the above identities, the coefficient of $(dR/dz)$ term in the radial Teukolsky equation reads, 
\begin{align}
\textrm{Coefficient~of}~\frac{dR}{dz}&=\frac{2iP_{+}-1}{z}-\frac{1+2iP_{+}+2i\omega\{2M+(r_{+}-r_{-})(1+z)\}}{(1+z)} \,,
\end{align}
which in the near-horizon approximation ($z\approx 0$), reduces to, 
\begin{align}
\textrm{Coefficient~of}~\frac{dR}{dz}\Bigg|_{\rm near~horizon}&=\frac{2iP_{+}-1}{z}-\frac{1+2iP_{+}+2i\omega\{2M+(r_{+}-r_{-})\}}{(1+z)}~.
\end{align}
Note that there are no terms of $\mathcal{O}(M^{2}\omega^{2})$ in the coefficient of $(dR/dz)$. Similarly, the coefficient of the radial Teukolsky function $R$, in the radial Teukolsky equation, becomes, 
\begin{align}
\textrm{Coefficient~of}~R&=-\frac{4iP_{+}}{z^{2}}+\frac{4iP_{+}+8iM\omega}{(1+z)^{2}}-\frac{A_{+}+iB_{+}}{z(1+z)}+\frac{(A_{-}-A_{+})+i(B_{-}-B_{+})}{(1+z)}
\nonumber
\\
&=-\frac{4iP_{+}}{z^{2}}+\frac{4iP_{+}+8iM\omega}{(1+z)^{2}}-\frac{\ell(\ell+1)-2+(E_{1}-2m)a\omega+2i\omega r_{+}+\omega^{2}(-r_{+}^{2}+2Mr_{+}+E_{2}a^{2})}{z(1+z)}
\nonumber
\\
&\qquad+\frac{\omega^{2}(r_{+}-r_{-})(r_{+}+r_{-}-2M)-2i\omega(r_{+}-r_{-})}{(1+z)}
\nonumber
\\
&=-\frac{4iP_{+}}{z^{2}}+\frac{4iP_{+}+2i\omega\{4M-(1+z)(r_{+}-r_{-})\}+\omega^{2}(r_{+}-r_{-})(r_{+}+r_{-}-2M)(1+z)}{(1+z)^{2}}
\nonumber
\\
&\qquad-\frac{\ell(\ell+1)-2}{z(1+z)}+\frac{2ma\omega}{z(1+z)}\left(1-\frac{E_{1}}{2m}\right)-\frac{2i\omega r_{+}}{z(1+z)}-\frac{\omega^{2}(-r_{+}^{2}+2Mr_{+}+E_{2}a^{2})}{z(1+z)}~,
\end{align}
which depends on terms involving $M^{2}\omega^{2}$. Using the result that $r_{+}+r_{-}=2M$, as well as $r_{+}-2M=-r_{-}$ and $r_{+}r_{-}=a^{2}$, along with the near-horizon approximation, the coefficient of the Teukolsky radial function in the Teukolsky equation becomes,
\begin{align}
\textrm{Coefficient~of}~R\Bigg|_{\rm near~horizon}&=-\frac{4iP_{+}}{z^{2}}+\frac{4iP_{+}+2i\omega\{4M-(r_{+}-r_{-})\}}{(1+z)^{2}}
\nonumber
\\
&\qquad-\frac{\ell(\ell+1)-2}{z(1+z)}+\frac{2ma\omega}{z(1+z)}\left(1-\frac{E_{1}}{2m}\right)-\frac{2i\omega r_{+}}{z(1+z)}-\frac{\omega^{2}a^{2}(1+E_{2})}{z(1+z)}~.
\end{align}
These expressions for the coefficients of $(dR/dz)$ and $R$ term in the radial Teukolsky equation have been used in the main text. 
\section{Zero-rotation limit of the Detweiler function}\label{zero_rot_detweiler}

In this appendix, we will discuss the perturbation of compact objects having zero rotation. In this limit, we first write down the relation between the Detweiler and the Teukolsky function with $s=-2$, which reads~\cite{Maggio:2019zyv}, 
\begin{align}
\,_{-2}X^{0}_{\ell m}&=\left\{\left(\frac{r}{r^{2}-2Mr}\right)\alpha_{-2}^{0}\right\}\,_{-2}R_{\ell m}^{0}+\left\{\left(\frac{r}{\left(r^{2}-2Mr\right)^{2}}\right)\beta_{-2}^{0}\right\}\dfrac{\,_{-2}R_{\ell m}^{0}}{dr}
\nonumber
\\
&=\left\{\left(\frac{1}{r-2M}\right)\alpha_{-2}^{0}\right\}\,_{-2}R_{\ell m}^{0}+\left\{\left(\frac{1}{\left(r-2M\right)^{3}}\right)\beta_{-2}^{0}\right\}\dfrac{\,_{-2}R_{\ell m}^{0}}{dr_{*}}\,,
\end{align}
where $\alpha_{-2}^{0}$ and $\beta_{-2}^{0}$ are the coefficients necessary to make the potential for the Detweiler function real with zero rotation and $r_{*}$ is the tortoise coordinate. For $s=-2$ and in the zero rotation limit, the functions $\alpha_{-2}^{0}$ and $\beta_{-2}^{0}$ read as follows~\cite{Maggio:2019zyv},
\begin{align}
\alpha_{-2}^{0}&=\frac{\kappa(A_{1}+iA_{2})+|\kappa|^{2}}{\sqrt{2}|\kappa|\sqrt{A_{1}+\textrm{Re} \kappa}}~,
\\
\beta_{-2}^{0}&=\frac{i\kappa B_{2}}{\sqrt{2}|\kappa|\sqrt{A_{1}+\textrm{Re} \kappa}}~,
\end{align}
where, we have introduced the quantities $\kappa$, $A_{1}$, $A_{2}$ and $B_{2}$, defined as,
\begin{align}
\kappa&=4\gamma_{\ell}(\gamma_{\ell}+2)+48i\omega M~,
\qquad
\gamma_{\ell}=\left(\ell+2\right)\left(\ell-1\right)~,
\\
A_{1}&=4\left[\frac{8r^{8}\omega^{4}}{(r^{2}-2Mr)^{2}}+\frac{8r^{4}\omega^{2}}{(r^{2}-2Mr)}\left(\frac{M^{2}}{(r^{2}-2Mr)}-\gamma_{\ell}\right)-\frac{4r^{2}\omega^{2}}{(r^{2}-2Mr)}\left(3r^{2}+2Mr\right)+12r^{2}\omega^{2}+\gamma_{\ell}(\gamma_{\ell}+2)\right]~,
\\
A_{2}&=4\left[-\frac{24r^{5}\omega^{3}}{(r^{2}-2Mr)}+\frac{4\gamma_{\ell}(r-M)r^{2}\omega}{(r^{2}-2Mr)}+4\omega r\gamma_{\ell}+12\omega M \right]~,
\\
B_{2}&=32r^{6}\omega^{3}+16r^{2}\omega(r^{2}-2Mr)\left\{\frac{2M^{2}}{(r^{2}-2Mr)}-\gamma_{\ell}\right\}-8M\omega r(r^{2}-2Mr)~.
\end{align}
Further, the Teukolsky function $\,_{-2}R_{\ell m}^{0}$ can be related to the Regge-Wheeler and  Zerilli function in the zero-rotation limit. The first step in this direction is to decompose the Teukolsky function into polar and axial parts, 
\begin{align}
\,_{-2}R_{\ell m}^{0}=\,_{-2}R_{\ell m}^{0,~\textrm{axial}}+\,_{-2}R_{\ell m}^{0,~\textrm{polar}}~.
\end{align}
Since the Detweiler function depends on the Teukolsky function in a linear manner, it follows that such a decomposition will hold for the Detweiler function as well,
\begin{align}
\,_{-2}X_{\ell m}^{0}=\,_{-2}X_{\ell m}^{0,~\textrm{axial}}+\,_{-2}X_{\ell m}^{0,~\textrm{polar}}~.
\end{align}
The axial and polar parts of the Teukolsky function can then be expressed in terms of the Regge-Wheeler and Zerilli functions, respectively, as
\begin{align}
\,_{-2}R_{\ell m}^{0,~\textrm{axial}}&=\frac{r^{3}\sqrt{\gamma_{\ell}(2+\gamma_{\ell})}}{8\omega}
\left[V^{\rm axial}\Psi^{\rm RW}_{\ell m}+\left(W^{\rm axial}+2i\omega\right)\left(\dfrac{d\Psi_{\ell m}^{\rm RW}}{dr_{*}}+i\omega \Psi_{\ell m}^{\rm RW}\right)\right]
\nonumber
\\
&=\frac{r^{3}\sqrt{\gamma_{\ell}(2+\gamma_{\ell})}}{8\omega}
\left[\left\{V^{\rm axial}+i\omega\left(W^{\rm axial}+2i\omega\right)\right\}\Psi^{\rm RW}_{\ell m}+\left(W^{\rm axial}+2i\omega\right)\left(\dfrac{d\Psi_{\ell m}^{\rm RW}}{dr_{*}}\right)\right]~,
\\
\,_{-2}R_{\ell m}^{0,~\textrm{polar}}&=-\frac{r^{3}\sqrt{\gamma_{\ell}(2+\gamma_{\ell})}}{8}
\left[\left\{V^{\rm polar}+i\omega\left(W^{\rm polar}+2i\omega\right)\right\}\Psi^{\rm Z}_{\ell m}+\left(W^{\rm polar}+2i\omega\right)\left(\dfrac{d\Psi^{\rm Z}_{\ell m}}{dr_{*}}\right)\right]~,
\end{align}
where we have introduced four functions of the radial coordinate $r$, namely $V_{\rm axial}$, $V_{\rm polar}$, $W_{\rm axial}$ and $W_{\rm polar}$, each of which can be defined as,
\begin{align}
V^{\rm axial}&=\frac{(r-2M)\left\{(\gamma_{\ell}+2)r-6M\right\}}{r^{4}}=V_{\rm RW}~,
\\
W^{\rm axial}&=\frac{2(r-3M)}{r^{2}}~,
\\
V^{\rm polar}&=\frac{(r-2M)}{r^{4}(\gamma_{\ell}r+6M)^{2}}\left[\gamma_{\ell}^{2}(\gamma_{\ell}+2)r^{3}+6M\gamma_{\ell}^{2}r^{2}+36M^{2}\gamma_{\ell}r+72M^{3}\right]=V_{\rm Z}~,
\\
W^{\rm polar}&=\frac{2\gamma_{\ell}r^{2}-6\gamma_{\ell}Mr-12M^{2}}{r^{2}(\gamma_{\ell}r+6M)}~.
\end{align}
Therefore, one can directly relate the axial and polar parts of the Detweiler function to the Regge-Wheeler and the Zerilli function, respectively,
\begin{align}
\,_{-2}X_{\ell m}^{\rm axial}&=\left(\frac{\alpha_{-2}^{0}}{r-2M}\right)\,_{-2}R_{\ell m}^{0,~\textrm{axial}}+\left(\frac{\beta_{-2}^{0}}{\left(r-2M\right)^{3}}\right)\dfrac{\,_{-2}R_{\ell m}^{0,~\textrm{axial}}}{dr_{*}}
\nonumber
\\
&=\left(\frac{\alpha_{-2}^{0}}{r-2M}\right)\Bigg[\frac{r^{3}\sqrt{\gamma_{\ell}(2+\gamma_{\ell})}}{8\omega}
\left[\left\{V^{\rm axial}+i\omega\left(W^{\rm axial}+2i\omega\right)\right\}\Psi^{\rm RW}_{\ell m}+\left(W^{\rm axial}+2i\omega\right)\left(\dfrac{d\Psi^{\rm RW}_{\ell m}}{dr_{*}}\right)\right]\Bigg]
\nonumber
\\
&+\left(\frac{\beta_{-2}^{0}}{\left(r-2M\right)^{3}}\right)\frac{d}{dr_{*}}\Bigg[\frac{r^{3}\sqrt{\gamma_{\ell}(2+\gamma_{\ell})}}{8\omega}
\left[\left\{V^{\rm axial}+i\omega\left(W^{\rm axial}+2i\omega\right)\right\}\Psi^{\rm RW}_{\ell m}+\left(W^{\rm axial}+2i\omega\right)\left(\dfrac{d\Psi^{\rm RW}_{\ell m}}{dr_{*}}\right)\right]\Bigg]
\nonumber
\\
&\equiv A_{\rm RW}(r)\Psi^{\rm RW}_{\ell m}+B_{\rm RW}(r)\left(\frac{\Psi^{\rm RW}_{\ell m}}{dr}\right)~,
\end{align}
where, we have defined the two functions $A_{\rm RW}$ and $B_{\rm RW}$, in the following expressions,
\begin{align}
A_{\rm RW}(r)&=\left(\frac{\alpha_{-2}^{0}}{r-2M}\right)\frac{r^{3}\sqrt{\gamma_{\ell}(2+\gamma_{\ell})}}{8\omega}\left\{V^{\rm axial}+i\omega\left(W^{\rm axial}+2i\omega\right)\right\}
\nonumber
\\
&\qquad +f(r)\left(\frac{\beta_{-2}^{0}}{\left(r-2M\right)^{3}}\right)\frac{d}{dr}\Bigg[\frac{r^{3}\sqrt{\gamma_{\ell}(2+\gamma_{\ell})}}{8\omega}
\left\{V^{\rm axial}+i\omega\left(W^{\rm axial}+2i\omega\right)\right\}\Bigg]
\nonumber
\\
&\qquad +\left(\frac{\beta_{-2}^{0}}{\left(r-2M\right)^{3}}\right)\left(\frac{r^{3}\sqrt{\gamma_{\ell}(2+\gamma_{\ell})}}{8\omega}\right)\left(W^{\rm axial}+2i\omega\right)V_{\rm RW}~,
\\
B_{\rm RW}(r)&=\left(\frac{\alpha_{-2}^{0}}{r-2M}\right)\frac{r^{3}\sqrt{\gamma_{\ell}(2+\gamma_{\ell})}}{8\omega}\left(W^{\rm axial}+2i\omega\right)f(r)
\nonumber
\\
&\qquad +f(r)\left(\frac{\beta_{-2}^{0}}{\left(r-2M\right)^{3}}\right)\Bigg[\frac{r^{3}\sqrt{\gamma_{\ell}(2+\gamma_{\ell})}}{8\omega}
\left\{V^{\rm axial}+i\omega\left(W^{\rm axial}+2i\omega\right)\right\}
\nonumber
\\
&\qquad\qquad +\frac{d}{dr}\left\{\frac{r^{3}\sqrt{\gamma_{\ell}(2+\gamma_{\ell})}}{8\omega}\left(W^{\rm axial}+2i\omega\right) \right\} \Bigg]~.
\end{align}
Hence, we can express, the ratio of the derivative of the axial part of the Detweiler function with the Detweiler function itself in terms of the same ratio for the Regge-Wheeler function as,
\begin{align}\label{DetRegr}
\frac{\left(\frac{\,_{-2}X_{\ell m}^{\rm axial}}{dr}\right)}{\,_{-2}X_{\ell m}^{\rm axial}}=\frac{\left[A_{\rm RW}'+B_{\rm RW}\left(\frac{r^{2}V_{\rm RW}}{(r-2M)^{2}}\right)\right]+\left[A_{\rm RW}+B_{\rm RW}'-B_{\rm RW}\left(\frac{2M}{r(r-2M)}\right)\right]\frac{\Psi_{\rm RW}'}{\Psi_{\rm RW}}}{A_{\rm RW}+B_{\rm RW}\frac{\Psi_{\rm RW}'}{\Psi_{\rm RW}}}~.
\end{align}
The above ratio can also be expressed in terms of the tortoise coordinate $r_{*}$,
\begin{align}\label{DetMetr}
\frac{1}{f}\frac{\left(\frac{\,_{-2}X_{\ell m}^{\rm axial}}{dr_{*}}\right)}{\,_{-2}X_{\ell m}^{\rm axial}}=\frac{f(r)\left[A_{\rm RW}'+B_{\rm RW}\left(\frac{r^{2}V_{\rm RW}}{(r-2M)^{2}}\right)\right]+\left[A_{\rm RW}+B_{\rm RW}'-B_{\rm RW}\left(\frac{2M}{r(r-2M)}\right)\right]\frac{\frac{d\Psi_{\rm RW}}{dr_{*}}}{\Psi_{\rm RW}}}{f(r)A_{\rm RW}+B_{\rm RW}\frac{\frac{d\Psi_{\rm RW}}{dr_{*}}}{\Psi_{\rm RW}}}~.
\end{align}
Along identical lines, for the polar sector, we obtain, 
\begin{align}
\,_{-2}X_{\ell m}^{\rm polar}=-A_{\rm Z}(r)\Psi^{\rm Z}_{\ell m}-B_{\rm Z}(r)\left(\frac{\Psi^{\rm Z}_{\ell m}}{dr}\right)~,
\end{align}
where the functions $A_{\rm Z}$ and $B_{\rm Z}$ are defined as,
\begin{align}
A_{\rm Z}(r)&=\left(\frac{\alpha_{-2}^{0}}{r-2M}\right)\frac{r^{3}\sqrt{\gamma_{\ell}(2+\gamma_{\ell})}}{8}\left\{V^{\rm polar}+i\omega\left(W^{\rm polar}+2i\omega\right)\right\}
\nonumber
\\
&\qquad +f(r)\left(\frac{\beta_{-2}^{0}}{\left(r-2M\right)^{3}}\right)\frac{d}{dr}\Bigg[\frac{r^{3}\sqrt{\gamma_{\ell}(2+\gamma_{\ell})}}{8}
\left\{V^{\rm polar}+i\omega\left(W^{\rm polar}+2i\omega\right)\right\}\Bigg]
\nonumber
\\
&\qquad +\left(\frac{\beta_{-2}^{0}}{\left(r-2M\right)^{3}}\right)\left(\frac{r^{3}\sqrt{\gamma_{\ell}(2+\gamma_{\ell})}}{8}\right)\left(W^{\rm polar}+2i\omega\right)V_{\rm Z}~,
\\
B_{\rm Z}(r)&=\left(\frac{\alpha_{-2}^{0}}{r-2M}\right)\frac{r^{3}\sqrt{\gamma_{\ell}(2+\gamma_{\ell})}}{8}\left(W^{\rm polar}+2i\omega\right)f(r)
\nonumber
\\
&\qquad +f(r)\left(\frac{\beta_{-2}^{0}}{\left(r-2M\right)^{3}}\right)\Bigg[\frac{r^{3}\sqrt{\gamma_{\ell}(2+\gamma_{\ell})}}{8}
\left\{V^{\rm polar}+i\omega\left(W^{\rm polar}+2i\omega\right)\right\}
\nonumber
\\
&\qquad\qquad +\frac{d}{dr}\left\{\frac{r^{3}\sqrt{\gamma_{\ell}(2+\gamma_{\ell})}}{8}\left(W^{\rm polar}+2i\omega\right) \right\} \Bigg]~.
\end{align}
Therefore, the ratio of the derivative of the polar sector of the Detweiler function with respect to the tortoise coordinate and the Detweiler function itself becomes,
\begin{align}
\frac{1}{f}\frac{\left(\frac{\,_{-2}X_{\ell m}^{\rm polar}}{dr_{*}}\right)}{\,_{-2}X_{\ell m}^{\rm polar}}=\frac{f(r)\left[A_{\rm Z}'+B_{\rm Z}\left(\frac{r^{2}V_{\rm Z}}{(r-2M)^{2}}\right)\right]+\left[A_{\rm Z}+B_{\rm Z}'-B_{\rm Z}\left(\frac{2M}{r(r-2M)}\right)\right]\frac{\frac{d\Psi_{\rm Z}}{dr_{*}}}{\Psi_{\rm Z}}}{f(r)A_{\rm Z}+B_{\rm Z}\frac{\frac{d\Psi_{\rm Z}}{dr_{*}}}{\Psi_{\rm Z}}}~.
\end{align}
These results have been used in the main text, as well as while discussing the static limit of the TLNs for non-rotating compact objects. 

\end{widetext}
\bibliography{References}

\bibliographystyle{./utphys1}
\end{document}